\documentclass[reqno,centertags, 12pt]{amsart}
\usepackage{amsmath,amsthm,amssymb,verbatim,graphicx}
\usepackage{latexsym}

\newcommand{\bbN}{{\mathbb{N}}}
\newcommand{\bbR}{{\mathbb{R}}}
\newcommand{\bbD}{{\mathbb{D}}}
\newcommand{\bbP}{{\mathbb{P}}}
\newcommand{\bbZ}{{\mathbb{Z}}}
\newcommand{\bbC}{{\mathbb{C}}}

\newcommand{\bbE}{{\mathbb{E}}}
\newcommand{\bbT}{{\mathbb{T}}}

\newcommand{\lb}{\label}

\newcommand{\bi}{\bibitem}

\newcommand{\beq}{\begin{equation}}
\newcommand{\eeq}{\end{equation}}
\newcommand{\ba}{\begin{align}}
\newcommand{\ea}{\end{align}}

\newcounter{smalllist}

\DeclareMathOperator{\Real}{Re}

\allowdisplaybreaks
\numberwithin{equation}{section}

\newtheorem{theorem}{Theorem}[section]

\newtheorem*{p2.1}{Proposition 2.1}

\newtheorem{lemma}[theorem]{Lemma}

\theoremstyle{definition}

\theoremstyle{remark}

\begin{document}
\title[The Statistical Distribution of the Zeros of Random OPUC]{The Statistical Distribution of the Zeros of Random Paraorthogonal Polynomials on the Unit Circle}
\author[Mihai Stoiciu]{Mihai Stoiciu}

\address{Mathematics 253-37, California Institute
of Technology, Pasadena, CA 91125.}
\email{mihai@its.caltech.edu}


\date{December 8, 2004}

\begin{abstract} We consider polynomials on the
unit circle defined by the recurrence relation
\[
\Phi_{k+1}(z) = z \Phi_{k} (z) - \overline{\alpha}_{k}
\Phi_k^{*}(z) \qquad k \geq 0, \quad \Phi_0=1
\]

For each $n$ we take $\alpha_0, \alpha_1, \ldots ,\alpha_{n-2}$
i.i.d. random variables distributed uniformly in a disk of radius $r
< 1$ and $\alpha_{n-1}$ another random variable independent of the
previous ones and distributed uniformly on the unit circle. The
previous recurrence relation gives a sequence of random
paraorthogonal polynomials $\{\Phi_n\}_{n \geq 0}$. For any $n$, the
zeros of $\Phi_n$ are $n$ random points on the unit circle.

We prove that for any $e^{i \theta} \in \partial \bbD$ the
distribution of the zeros of $\Phi_n$ in intervals of size
$O(\frac{1}{n})$ near $e^{i \theta}$ is the same as the distribution
of $n$ independent random points uniformly distributed on the unit
circle (i.e., Poisson). This means that, for large $n$, there is no
local correlation between the zeros of the considered random
paraorthogonal polynomials.
\end{abstract}

\maketitle

\section{Introduction} \lb{s1} In this paper we study the statistical
distribution of the zeros of paraorthogonal polynomials on the unit
circle. In order to introduce and motivate these polynomials, we
will first review a few aspects of the standard theory. Complete
references for both the classical and the spectral theory of
orthogonal polynomials on the unit circle are Simon \cite{OPUC1} and
\cite{OPUC2}.

One of the central results in this theory is Verblunsky's theorem,
which states that there is a one-one and onto map $\mu \rightarrow
\{\alpha_n\}_{n \geq 0}$ from the set of nontrivial (i.e., not
supported on a finite set) probability measures on the unit circle
and sequence of complex numbers $\{\alpha_n\}_{n \geq 0}$ with
$|\alpha_n| < 1$ for any $n$. The correspondence is given by the
recurrence relation obeyed by orthogonal polynomials on the unit
circle. Thus, if we apply the Gram-Schmidt procedure to the sequence
of polynomials $1, z, z^2, \ldots \in L^2(\partial \bbD, d \mu)$,
the polynomials obtained $\Phi_0(z, d\mu)$, $\Phi_1(z, d\mu)$,
$\Phi_2(z, d\mu) \ldots$ obey the recurrence relation
\begin{equation} \lb{polyrec}
\Phi_{k+1}(z,d\mu) = z \Phi_k(z,d\mu) - \overline{\alpha}_k
\Phi_k^{*} (z,d\mu) \qquad k \geq 0
\end{equation}
where for $\Phi_k(z,d\mu) = \sum_{j = 0}^k b_j z^j$, the reversed
polynomial $\Phi_k^{*}$ is given by $\Phi_k^{*} (z,d\mu) = \sum_{j
= 0}^k \overline{b}_{k-j} z^j$. The numbers $\alpha_k$ from
(\ref{polyrec}) obey $|\alpha_k| < 1$ and, for any $k$, the zeros
of the polynomial $\Phi_{k+1}(z,d\mu)$ lie inside the unit disk.

If, for a fixed $n$, we take $\alpha_0, \alpha_1, \ldots,
\alpha_{n-2} \in \bbD$ and $\alpha_{n-1} = \beta \in
\partial \bbD$ and we use the recurrence relations
$(\ref{polyrec})$ to define the polynomials $\Phi_0, \Phi_1, \ldots,
\Phi_{n-1}$, then the zeros of the polynomial
\begin{equation} \lb{l2.1.1}
\Phi_{n}(z,d\mu,\beta) = z \Phi_{n-1}(z,d\mu) - \overline{\beta}\,
\Phi_{n-1}^{*} (z,d\mu)
\end{equation}
are simple and situated on the unit circle. These polynomials
(obtained by taking the last Verblunsky coefficient on the unit
circle) are called paraorthogonal polynomials and were analyzed in
\cite{Go}, \cite{JNT}; see also Chapter 2 in Simon \cite{OPUC1}.

For any $n$, we will consider random Verblunsky coefficients by
taking $\alpha_0, \alpha_1, \ldots, \alpha_{n-2}$ to be i.i.d.
random variables distributed uniformly in a disk of fixed radius $r
< 1$ and $\alpha_{n-1}$ another random variable independent of the
previous ones and distributed uniformly on the unit circle.
Following the procedure mentioned before, we will get a sequence of
random paraorthogonal polynomials $\{\Phi_n =
\Phi_{n}(z,d\mu,\beta)\}_{n \geq 0}$. For any $n$, the zeros of
$\Phi_n$ are $n$ random points on the unit circle. Let us consider
\begin{equation} \lb{pointprocess}
\zeta^{(n)} = \sum_{k=1}^{n} \delta_{z_k^{(n)}}
\end{equation}
where $z_1^{(n)}, z_2^{(n)}, \ldots, z_n^{(n)}$ are the zeros of the
polynomial $\Phi_n$. Let us also fix a point $e^{i \theta} \in
\bbD$. We will prove that the distribution of the zeros of $\Phi_n$
on intervals of length $O(\frac{1}{n})$ situated near $e^{i \theta}$
is the same as the distribution of $n$ independent random points
uniformly distributed in the unit circle (i.e., Poisson).


A collection of random points on the unit circle is sometimes called
a point process on the unit circle. Therefore a reformulation of
this problem can be: The limit of the sequence point process
$\{\zeta_n\}_{n \geq 0}$ on a fine scale (of order $O(\frac{1}{n})$)
near a point $e^{i \theta}$ is a Poisson point process. This result
is illustrated by the following generic plot of the zeros of random
paraorthogonal polynomials:
\begin{center}
\includegraphics[scale=.5,angle=0]{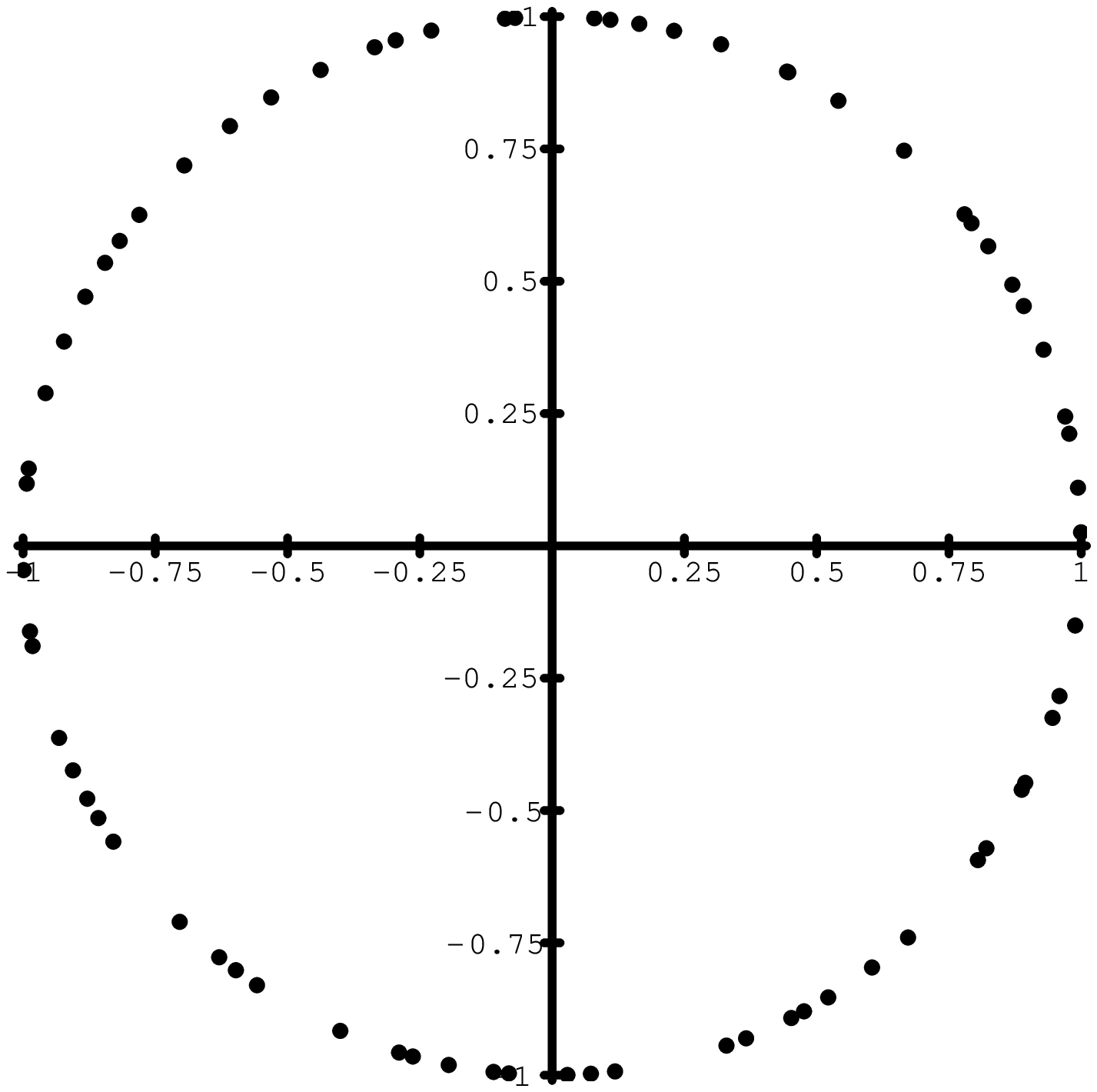}
\end{center}

This Mathematica plot represents the zeros of a paraorthogonal
polynomial of degree 71 obtained by randomly taking $\alpha_0,
\alpha_1, \ldots, \alpha_{69}$ from the uniform distribution on the
disk centered at the origin of radius $\frac{1}{2}$ and random
$\alpha_{70}$ from the uniform distribution on the unit circle. On a
fine scale we can observe some clumps, which suggests the Poisson
distribution.

Similar results appeared in the mathematical literature for the case
of random Schr\"odinger operators; see Molchanov \cite{Mo} and
Minami \cite{Mi}. The study of the spectrum of random Schr\"odinger
operators and of the distribution of the eigenvalues was initiated
by the very important paper of Anderson \cite{An}, who showed that
certain random lattices exhibit absence of diffusion. Rigorous
mathematical proofs of the Anderson localization were given by
Goldsheid-Molchanov-Pastur \cite{GMP} for one-dimensional models and
by Fr\"ohlich-Spencer \cite{FS} for multidimensional Schr\"odinger
operators. Several other proofs, containing improvements and
simplifications, were published later. We will only mention here
Aizenman-Molchanov \cite{AM} and Simon-Wolff \cite{SW}, which are
relevant for our approach. In the case of the unit circle, similar
localization results were obtained by Teplyaev \cite{Tep} and by
Golinskii-Nevai \cite{GN}.

In addition to the phenomenon of localization, one can also analyze
the local structure of the spectrum. It turns out that there is no
repulsion between the energy levels of the Schr\"odinger operator.
This was shown by Molchanov \cite{Mo} for a model of the
one-dimensional Schr\"odinger operator studied by the Russian
school. The case of the multidimensional discrete Schr\"odinger
operator was analyzed by Minami in \cite{Mi}. In both cases the
authors proved that the statistical distribution of the eigenvalues
converges locally to a stationary Poisson point process. This means
that there is no correlation between eigenvalues.

We will prove a similar result on the unit circle. For any
probability measure $d \mu$ on the unit circle, we denote by
$\{\chi_0, \chi_1, \chi_2, \ldots \}$ the basis of $L^2(\partial
\bbD, d \mu)$ obtained from $\{1, z, z^{-1}, z^2, z^{-2}, \ldots \}$
by applying the Gram-Schmidt procedure. The matrix representation of
the operator $f(z) \to z f(z)$ on $L^2(\partial \bbD, d \mu)$ with
respect to the basis $\{\chi_0, \chi_1, \chi_2, \ldots \}$ is a
five-diagonal matrix of the form: \begin{equation} \mathcal{C} =
\left(%
\begin{array}{cccccc}
  \bar{\alpha}_0 & \bar{\alpha}_1 \rho_0 & \rho_1 \rho_0 & 0 & 0 & \ldots \\
  \rho_0 & -\bar{\alpha}_1 \alpha_0 & -\rho_1 \alpha_0 & 0 & 0 & \ldots\\
  0 & \bar{\alpha}_2 \rho_1 & -\bar{\alpha}_2 \alpha_1 & \bar{\alpha}_3 \rho_2 & \rho_3 \rho_2 & \ldots\\
  0 & \rho_2 \rho_1 & -\rho_2 \alpha_1 & -\bar{\alpha}_3 \alpha_2 & -\rho_3 \alpha_2 & \ldots\\
  0 & 0 & 0 & \bar{\alpha}_4 \rho_3 & -\bar{\alpha}_4 \alpha_3 & \ldots\\
  \ldots & \ldots & \ldots & \ldots & \ldots &\ldots\\
\end{array}%
\right)
\end{equation}
($\alpha_0, \alpha_1, \ldots$ are the Verblunsky coefficients
associated to the measure $\mu$, and for any $n \geq 0$, $\rho_n =
\sqrt{1 - |\alpha_n|^2}$). This matrix representation is a recent
discovery of Cantero, Moral, and Vel\'asquez \cite{CMV}. The matrix
$\mathcal{C}$ is called the CMV matrix and will be used in the study
of the distribution of the zeros of the paraorthogonal polynomials.

Notice that if one of the $\alpha$'s is of absolute value 1, then
the Gram-Schmidt process ends and the CMV matrix decouples. In our
case, $|\alpha_{n-1}| = 1$, so $\rho_{n-1} = 0$ and therefore the
CMV matrix decouples between $(n-1)$ and $n$ and the upper left
corner is an $(n \times n)$ unitary matrix $\mathcal{C}^{(n)}$. The
advantage of considering this matrix is that the zeros of $\Phi_n$
are exactly the eigenvalues of the matrix $\mathcal{C}^{(n)}$ (see,
e.g., Simon \cite{OPUC1}). We will use some techniques from the
spectral theory of the discrete Schr\"odinger operators to study the
distribution of these eigenvalues, especially ideas and methods
developed in \cite{AM}, \cite{Aiz_et_al}, \cite{delRio}, \cite{Mi},
\cite{Mo}, \cite{SVan}. However, our model on the unit circle has
many different features compared to the discrete Schr\"odinger
operator (perhaps the most important one is that we have to consider
unitary operators on the unit circle instead of self-adjoint
operators on the real line). Therefore, we will have to use new
ideas and techniques that work for this situation.

The final goal is the following:

\begin{theorem} \lb{main} Consider the random polynomials on the unit circle given by the
following recurrence relations:
\begin{equation}
\Phi_{k+1}(z) = z \Phi_{k} (z) - \overline{\alpha}_{k} \Phi_k^{*}(z)
\qquad k \geq 0, \qquad \Phi_0=1 \end{equation} where $\alpha_0,
\alpha_1, \ldots, \alpha_{n-2}$ are i.i.d. random variables
distributed uniformly in a disk of radius $r < 1$ and $\alpha_{n-1}$
is another random variable independent of the previous ones and
uniformly distributed on the unit circle.

Consider the space \ $\Omega = \{ \alpha = (\alpha_0, \alpha_1,
\ldots, \alpha_{n-2}, \alpha_{n-1}) \in D(0,r) \times D(0,r) \times
\cdots \times D(0,r) \times
\partial \bbD \}$ with the probability measure $\bbP$ obtained by taking
the product of the uniform {\rm(}Lebesgue{\rm)} measures on each
$D(0,r)$ and on $\partial \bbD$. Fix a point $e^{i \theta_0} \in
\bbD$ and let $\zeta^{(n)}$ be the point process defined by
{\rm(}\ref{pointprocess}{\rm)}.

Then, on a fine scale {\rm(}of order $\frac{1}{n}${\rm)} near $e^{i
\theta_0}$, the point process $\zeta^{(n)}$ converges to the Poisson
point process with intensity measure $n \, \frac{d \theta}{2 \pi}$
{\rm (}where $\frac{d \theta}{2 \pi}$ is the normalized Lebesgue
measure{\rm)}. This means that for any fixed $a_1 < b_1 \leq a_2 <
b_2 \leq \cdots \leq a_m < b_m$ and any nonnegative integers $k_1,
k_2, \ldots, k_m$, we have:
\begin{align} \lb{CMT}
\bbP \left( \zeta^{(n)} \left( e^{i (\theta_0 + \frac{2 \pi
a_1}{n})}, e^{i (\theta_0 + \frac{2 \pi b_1}{n})}\right) = k_1,
\ldots, \ \zeta^{(n)} \left( e^{i (\theta_0 + \frac{2 \pi a_m}{n})},
e^{i (\theta_0 + \frac{2 \pi b_m}{n})}\right) = k_m
\right) \notag\\
\longrightarrow e^{-(b_1 - a_1)} \frac{(b_1 - a_1)^{k_1}}{k_1!}
\ldots e^{-(b_m - a_m)} \frac{(b_m - a_m)^{k_m}}{k_m !}
\end{align} as $n \to \infty$.

\end{theorem}

\section{Outline of the Proof} \lb{s2}

From now on we will work under the hypotheses of Theorem~\ref{main}.
We will study the statistical distribution of the eigenvalues of the
random CMV matrices
\begin{equation} \lb{model}
\mathcal{C}^{(n)} = \mathcal{C}^{(n)}_{\alpha}
\end{equation}
for $\alpha \in \Omega$ (with the space $\Omega$ defined in
Theorem~\ref{main}).

A first step in the study of the spectrum of random CMV matrix is
proving the exponential decay of the fractional moments of the
resolvent of the CMV matrix. These ideas were developed in the case
of Anderson models by Aizenman-Molchanov \cite{AM} and by Aizenman
et al. \cite{Aiz_et_al}. In the case of Anderson models, they
provide a powerful method for proving spectral localization,
dynamical localization, and the absence of level repulsion.

Before we state the Aizenman-Molchanov bounds, we have to make a few
remarks on the boundary behavior of the matrix elements of the
resolvent of the CMV matrix. For any $z \in \bbD$ and any $0 \leq
k,l \leq (n-1)$, we will use the following notation:
\begin{equation}
F_{k l}(z, \mathcal{C}^{(n)}_{\alpha}) =
\left[\frac{\mathcal{C}^{(n)}_{\alpha} +
z}{\mathcal{C}^{(n)}_{\alpha} - z}\right]_{kl}
\end{equation}

As we will see in the next section, using properties of
Carath\'eodory functions, we will get that for any $\alpha \in
\Omega$, the radial limit
\begin{equation}
F_{k l} (e^{i \theta}, \mathcal{C}^{(n)}_{\alpha}) = \lim_{r
\uparrow 1} F_{k l} (r e^{i \theta}, \mathcal{C}^{(n)}_{\alpha})
\end{equation} exists for Lebesgue almost every $e^{i \theta} \in
\partial \bbD$ and $F_{k l} (\,\cdot\,, \mathcal{C}^{(n)}_{\alpha})
\in L^s (\partial \bbD)$ for any $s \in (0,1)$. Since the
distributions of $\alpha_0, \alpha_1, \ldots, \alpha_{n-1}$ are
rotationally invariant, we obtain that for any fixed $e^{i \theta}
\in \partial \bbD$, the radial limit $F_{k l} (e^{i \theta},
\mathcal{C}^{(n)}_{\alpha})$ exists for almost every $\alpha \in
\Omega$. We can also define
\begin{align}
G_{k l}(z, \mathcal{C}^{(n)}_{\alpha}) &=
\left[\frac{1}{\mathcal{C}^{(n)}_{\alpha} - z}\right]_{kl}
\intertext{and} G_{k l} (e^{i \theta}, \mathcal{C}^{(n)}_{\alpha})
&= \lim_{r \uparrow 1} G_{k l} (r e^{i \theta},
\mathcal{C}^{(n)}_{\alpha})
\end{align}

Using the previous notation we have:

\medskip

\begin{theorem}[Aizenman-Molchanov Bounds for the Resolvent of the CMV Matrix] \lb{t1} For the model considered in Theorem~\ref{main} and for any $s \in (0,1)$, there exist
constants $C_1, D_1
> 0$ such that for any $n > 0$, any $k,l$, $0 \leq k, l \leq n-1$ and any $e^{i \theta} \in \partial \bbD$, we have:
\begin{equation} \lb{expdec}
\bbE \left(\left| F_{k l} (e^{i \theta}, \mathcal{C}^{(n)}_{\alpha})
\right|^s\right) \leq C_1 \, e^{- D_1 |k - l|}
\end{equation} where $\mathcal{C}^{(n)}$ is the $(n \times n)$ CMV
matrix obtained for $\alpha_0, \alpha_1, \ldots \alpha_{n-2}$
uniformly distributed in $D(0,r)$ and $\alpha_{n-1}$ uniformly
distributed in $\partial \bbD$.
\end{theorem}

\medskip

Using Theorem \ref{t1}, we will then be able to control the
structure of the eigenfunctions of the matrix $\mathcal{C}^{(n)}$.

\medskip

\begin{theorem}[The Localized Structure of the Eigenfunctions] \lb{t2} For
the model considered in Theorem \ref{main}, the eigenfunctions of
the random matrices $\mathcal{C}^{(n)} = \mathcal{C}^{(n)}_{\alpha}$
are exponentially localized with probability $1$, that is
exponentially small outside sets of size proportional to $(\ln n)$.
This means that there exists a constant $D_2 > 0$ and for almost
every $\alpha \in \Omega$, there exists a constant $C_{\alpha} > 0$
such that for any unitary eigenfunction $\varphi_{\alpha}^{(n)}$,
there exists a point $m({\varphi_{\alpha}^{(n)}})$ $(1 \leq
m({\varphi_{\alpha}^{(n)}}) \leq n)$ with the property that for any
$m, |m - m({\varphi_{\alpha}^{(n)}})| \geq D_2 \ln (n+1)$, we have
\begin{equation}
|\varphi_{\alpha}^{(n)}(m)| \leq C_{\alpha} \, e^{- (4/D_2) \, |m \,
- \, m({\varphi_{\alpha}^{(n)}})|}
\end{equation}
\end{theorem}

The point $m({\varphi_{\alpha}^{(n)}})$ will be taken to be the
smallest integer where the eigenfunction
$\varphi_{\alpha}^{(n)}(m)$ attains its maximum.

\bigskip

In order to obtain a Poisson distribution in the limit as $n \to
\infty$, we will use the approach of Molchanov \cite {Mo} and Minami
\cite{Mi}. The first step is to decouple the point process
$\zeta^{(n)}$ into the direct sum of smaller point processes. We
will do the decoupling process in the following way: For any
positive integer $n$, let $\tilde{\mathcal{C}}^{(n)}$ be the CMV
matrix obtained for the coefficients $\alpha_0, \alpha_1, \ldots,
\alpha_n$ with the additional restrictions $\alpha_{\left[
\frac{n}{\ln n}\right]} = e^{i \eta_1}$, $\alpha_{2 \left[
\frac{n}{\ln n} \right]} = e^{i \eta_2}$, \ldots,  $\alpha_n = e^{i
\eta_{[\ln n]}}$, where $e^{i \eta_1}, e^{i \eta_2}, \ldots, e^{i
\eta_{[\ln n]}}$ are independent random points uniformly distributed
on the unit circle. Note that the matrix $\tilde{\mathcal{C}}^{(n)}$
decouples into the direct sum of \,$\approx [\ln n]$ unitary
matrices $\tilde{\mathcal{C}}^{(n)}_1$,
$\tilde{\mathcal{C}}^{(n)}_2$, \ldots,
$\tilde{\mathcal{C}}^{(n)}_{\left[ \ln n \right]}$. We should note
here that the actual number of blocks $\tilde{\mathcal{C}}^{(n)}_i$
is slightly larger than $[\ln n]$ and that the dimension of one of
the blocks (the last one) could be smaller than $\left[\frac{n}{\ln
n}\right]$.

However, since we are only interested in the asymptotic behavior of
the distribution of the eigenvalues, we can, without loss of
generality, work with matrices of size $N = \left[ \ln n \right]
\left[ \frac{n}{\ln n} \right]$. The matrix
$\tilde{\mathcal{C}}^{(N)}$ is the direct sum of exactly $[\ln n]$
smaller blocks $\tilde{\mathcal{C}}^{(N)}_1$,
$\tilde{\mathcal{C}}^{(N)}_2$, \ldots,
$\tilde{\mathcal{C}}^{(N)}_{\left[ \ln n \right]}$. We denote by
$\zeta^{(N,p)} = \sum_{k=1}^{\left[n/\ln n\right]}
\delta_{z_k^{(p)}}$ where $z_1^{(p)}, z_2^{(p)}, \ldots, z_{\left[n/
\ln n\right]}^{(p)}$ are the eigenvalues of the matrix
$\tilde{\mathcal{C}}^{(N)}_p$. The decoupling result is formulated
in the following theorem:

\begin{theorem}[Decoupling the point process] \lb{t3} The
point process $\zeta^{(N)}$ can be asymptotically approximated by
the direct sum of point processes $\sum_{p = 1}^{\left[ \ln n
\right]} \zeta^{(N,p)}$. In other words, the distribution of the
eigenvalues of the matrix $\mathcal{C}^{(N)}$ can be asymptotically
approximated by the distribution of the eigenvalues of the direct
sum of the matrices $\tilde{\mathcal{C}}^{(N)}_1$,
$\tilde{\mathcal{C}}^{(N)}_2$, \ldots,
$\tilde{\mathcal{C}}^{(N)}_{\left[ \ln n \right]}$.
\end{theorem}

The decoupling property is the first step in proving that the
statistical distribution of the eigenvalues of $\mathcal{C}^{(N)}$
is Poisson. In the theory of point processes (see, e.g., Daley and
Vere-Jones \cite{DV}), a point process obeying this decoupling
property is called an infinitely divisible point process. In order
to show that this distribution is Poisson on a scale of order
$O(\frac{1}{n})$ near a point $e^{i \theta}$, we need to check two
conditions:

\begin{align}
& \textrm{i)} \quad \sum_{p = 1}^{\left[ \ln n \right]} \bbP \left(
\zeta^{(N,p)}
\left(A\left(N, \theta\right)\right) \geq 1 \right) \rightarrow |A| \quad \textrm{as} \quad n \to \infty\\
&\textrm{ii)} \quad \sum_{p = 1}^{\left[ \ln n \right]} \bbP \left(
\zeta^{(N,p)}\left(A \left(N,\theta \right)\right) \geq 2 \right)
\rightarrow 0 \quad \textrm{as} \quad n \to \infty
\end{align}
where for an interval $A = [a,b]$ we denote by $A(N, \theta) = (e^{i
(\theta + \frac{2 \pi a}{N})}, e^{i (\theta + \frac{2 \pi b}{N})})$
and $|\,\cdot\,|$ is the Lebesgue measure (and we extend this
definition to unions of intervals). The second condition shows that
it is asymptotically impossible that any of the matrices
$\tilde{\mathcal{C}}^{(N)}_1$, $\tilde{\mathcal{C}}^{(N)}_2$,
\ldots, $\tilde{\mathcal{C}}^{(N)}_{\left[ \ln n \right]}$ has two
or more eigenvalues situated an interval of size $\frac{1}{N}$.
Therefore, each of the matrices $\tilde{\mathcal{C}}^{(N)}_1$,
$\tilde{\mathcal{C}}^{(N)}_2$, \ldots,
$\tilde{\mathcal{C}}^{(N)}_{\left[ \ln n \right]}$ contributes with
at most one eigenvalue in an interval of size $\frac{1}{N}$. But the
matrices $\tilde{\mathcal{C}}^{(N)}_1$,
$\tilde{\mathcal{C}}^{(N)}_2$, \ldots,
$\tilde{\mathcal{C}}^{(N)}_{\left[ \ln n \right]}$ are decoupled,
hence independent, and therefore we get a Poisson distribution. The
condition $\rm{i)}$ now gives Theorem~\ref{main}.

\medskip

The next four sections will contain the detailed proofs of these
theorems.


\bigskip

\section{Aizenman-Molchanov Bounds for the Resolvent of the CMV Matrix} \lb{s3}

We will study the random CMV matrices defined in (\ref{model}). We
will analyze the matrix elements of the resolvent
$(\mathcal{C}^{(n)} - z)^{-1}$ of the CMV matrix, or, what is
equivalent, the matrix elements of
\begin{equation}
F(z,\mathcal{C}^{(n)}) = (\mathcal{C}^{(n)} + z)(\mathcal{C}^{(n)} -
z)^{-1} = I + 2z \, (\mathcal{C}^{(n)} - z)^{-1}
\end{equation} (we consider $z \in \bbD$). More precisely, we will
be interested in the expectations of the fractional moments of
matrix elements of the resolvent. This method (sometimes called the
fractional moments method) is useful in the study of the eigenvalues
and of the eigenfunctions and was introduced by Aizenman and
Molchanov in \cite{AM}.

We will prove that the expected value of the fractional moment of
the matrix elements of the resolvent decays exponentially (see
(\ref{expdec})). The proof of this result is rather involved; the
main steps will be:

Step 1. The fractional moments $\bbE\left(\left| F_{k l} (z,
\mathcal{C}^{(n)}_{\alpha}) \right|^s\right)$ are uniformly bounded
(Lemma \ref{l1}).

Step 2. The fractional moments $\bbE\left(\left| F_{k l} (z,
\mathcal{C}^{(n)}_{\alpha}) \right|^s\right)$ converge to 0
uniformly along the rows (Lemma \ref{unifdec}).

Step 3. The fractional moments $\bbE\left(\left| F_{k l} (z,
\mathcal{C}^{(n)}_{\alpha}) \right|^s\right)$ decay exponentially
(Theorem \ref{t1}).

\medskip

We will now begin the analysis of $\bbE\left(\left| F_{k l} (z,
\mathcal{C}^{(n)}_{\alpha}) \right|^s\right)$.

It is not hard to see that $\Real \left[(\mathcal{C}^{(n)} +
z)(\mathcal{C}^{(n)} - z)^{-1}\right]$ is a positive operator.
This will help us prove:

\begin{lemma} \lb{l1} For any $s
\in (0,1)$, any $k,l$, $1 \leq k,l \leq n$, and any $z \in \bbD \cup
\partial \bbD$, we have
\begin{equation}
\bbE\left(\left| F_{k l} (z, \mathcal{C}^{(n)}_{\alpha})
\right|^s\right) \leq C
\end{equation} where $C=\frac{2^{2-s}}{\cos
\frac{\pi s}{2}}$.
\end{lemma}

\begin{proof}
Let $F_{\varphi}(z) = (\varphi, (\mathcal{C}^{(n)}_{\alpha} +
z)(\mathcal{C}^{(n)}_{\alpha} - z)^{-1} \varphi)$. Since $\Real
F_{\varphi} \geq 0$, the function $F_{\varphi}$ is a Carath\'eodory
function for any unit vector $\varphi$. Fix $\rho \in (0,1)$. Then,
by a version of Kolmogorov's theorem (see Duren \cite{Duren} or
Khodakovsky \cite{Kho}),
\begin{equation} \lb{b0}
\int_{0}^{2 \pi} \left|(\varphi, (\mathcal{C}^{(n)}_{\alpha} + \rho
e^{i \theta})(\mathcal{C}^{(n)}_{\alpha} - \rho e^{i \theta})^{-1}
\varphi)\right|^s \frac{d\theta}{2 \pi} \, \leq \, C_1
\end{equation} where $C_1 = \frac{1}{\cos \frac{\pi s}{2}}$.

The polarization identity gives (assuming that our scalar product is
antilinear in the first variable and linear in the second variable)
\begin{equation} \lb{linCa}
F_{k l} (\rho e^{i \theta}, \mathcal{C}^{(n)}_{\alpha}) =
\frac{1}{4} \, \sum_{m = 0}^{3} (-i)^m \left( (\delta_k + i^m
\delta_l),\, F(\rho e^{i \theta}, \mathcal{C}^{(n)}_{\alpha})
(\delta_k + i^m \delta_l) \right)
\end{equation}
which, using the fact that $|a+b|^s \leq |a|^s + |b|^s$, implies
\begin{align}
\left|F_{k l} (\rho e^{i \theta},
\mathcal{C}^{(n)}_{\alpha})\right|^s \leq \frac{1}{2^s} \, \sum_{m =
0}^{3} \left|\left( \frac{(\delta_k + i^m \delta_l)}{\sqrt{2}},\,
F(\rho e^{i \theta}, \mathcal{C}^{(n)}_{\alpha}) \frac{(\delta_k +
i^m \delta_l)}{\sqrt{2}} \right)\right|^s \lb{bound3.5}
\end{align}

Using (\ref{b0}) and (\ref{bound3.5}), we get, for any
$\mathcal{C}^{(n)}_{\alpha}$,
\begin{equation}
\int_0^{2 \pi} \left| F_{k l} (\rho e^{i \theta},
\mathcal{C}^{(n)}_{\alpha}) \right|^s \, \frac{d\theta}{2 \pi} \leq
C
\end{equation}
where $C = \frac{2^{2-s}}{\cos \frac{\pi s}{2}}$.

Therefore, after taking expectations and using Fubini's theorem,
\begin{equation} \lb{b}
\int_{0}^{2 \pi} \bbE \left(\left|F_{k l} (\rho e^{i \theta},
\mathcal{C}^{(n)}_{\alpha})\right|^s \right) \frac{d\theta}{2 \pi}
\, \leq C
\end{equation}

The coefficients $\alpha_0, \alpha_1, \ldots, \alpha_{n-1}$ define a
measure $d \mu$ on $\partial \bbD$. Let us consider another measure
$d \mu_{\theta} (e^{i \tau}) = d \mu (e^{i (\tau - \theta)})$. This
measure defines Verblunsky coefficients $\alpha_{0, \theta},
\alpha_{1, \theta}, \ldots, \alpha_{n-1, \theta}$, a CMV matrix
$\mathcal{C}^{(n)}_{\alpha,\theta}$, and unitary orthogonal
polynomials $\varphi_{0, \theta}, \varphi_{1, \theta} \ldots,
\varphi_{n-1, \theta}$. Using the results presented in Simon
\cite{OPUC1}, for any $k$, $0 \leq k \leq n-1$,
\begin{align}
\alpha_{k, \theta} &= e^{- i (k+1) \theta} \alpha_k \lb{b5}\\
\varphi_{k, \theta} (z) &= e^{i k \theta} \varphi_{k} (e^{-i \theta}
z) \lb{b6}
\end{align}

The relation (\ref{b6}) shows that for any $k$ and $\theta$,
$\chi_{k, \theta} (z) = \lambda_{k,\theta} \, \chi_{k} (e^{-i
\theta} z)$ where $|\lambda_{k, \theta}| = 1$.

Since $\alpha_0, \alpha_1, \ldots, \alpha_{n-1}$ are independent and
the distribution of each one of them is rotationally invariant, we
have
\begin{equation}
\bbE \left( \left| F_{k l} (\rho e^{i \theta},
\mathcal{C}^{(n)}_{\alpha}) \right|^s \right) = \bbE
\left(\left|F_{k l} (\rho e^{i \theta},
\mathcal{C}^{(n)}_{\alpha,\theta})\right|^s \right)
\end{equation}

But, using (\ref{b5}) and (\ref{b6}),
\begin{align*}
F_{k l} (\rho e^{i \theta}, \mathcal{C}^{(n)}_{\alpha,\theta}) &=
\int_{\partial \bbD} \frac{e^{i \tau} + \rho e^{i \theta}}{e^{i
\tau}-\rho e^{i \theta}} \, \chi_{l, \theta} (e^{i \tau}) \,
\overline{\chi_{k, \theta} (e^{i
\tau})} \, d \mu_{\theta} (e^{i \tau})\\
&=\int_{\partial \bbD} \frac{e^{i \tau} + \rho e^{i \theta}}{e^{i
\tau}-\rho e^{i \theta}} \, \chi_{l, \theta} (e^{i \tau}) \,
\overline{\chi_{k, \theta} (e^{i \tau})} \, d \mu (e^{i (\tau -
\theta)})\\ &= \int_{\partial \bbD} \frac{e^{i (\tau + \theta)} +
\rho e^{i \theta}}{e^{i (\tau + \theta)}-\rho e^{i \theta}} \,
\chi_{l, \theta} (e^{i (\tau + \theta)}) \, \overline{\chi_{k,
\theta} (e^{i (\tau + \theta)})} \, d \mu
(e^{i \tau})\\
&= \lambda_{l,\theta} \overline{\lambda}_{k,\theta} \int_{\partial
\bbD} \frac{e^{i \tau} + \rho}{e^{i \tau}-\rho} \, \chi_{l} (e^{i
\tau}) \,
\overline{\chi_{k} (e^{i \tau})} \, d \mu (e^{i \tau})\\
&= \lambda_{l,\theta} \, \overline{\lambda}_{k,\theta} \, F_{k l}
(\rho, \mathcal{C}^{(n)}_{\alpha})
\end{align*}
where $|\lambda_{l,\theta} \overline{\lambda}_{k,\theta}| = 1$.

Therefore the function $\theta \rightarrow \bbE \left(\left|F_{k l}
(\rho e^{i \theta}, \mathcal{C}^{(n)}_{\alpha})\right|^s \right)$ is
constant, so, using (\ref{b}), we get
\begin{equation} \lb{b7}
\bbE \left(\left|F_{k l} (\rho e^{i \theta},
\mathcal{C}^{(n)}_{\alpha,\theta})\right|^s \right) \leq C
\end{equation}
Since $\rho$ and $\theta$ are arbitrary, we now get the desired
conclusion for any $z \in \bbD$.

Observe that, by (\ref{linCa}), $F_{k l}$ is a linear combination of
Carath\'eodory functions. By \cite{Duren}, any Carath\'eodory
function is in $H^s(\bbD)$ ($0 < s < 1$) and therefore it has
boundary values almost everywhere on $\partial \bbD$. Thus we get
that, for any fixed $\alpha \in \Omega$ and for Lebesgue almost any
$z = e^{i \theta} \in \partial \bbD$, the radial limit $F_{k l}
(e^{i \theta}, \mathcal{C}^{(n)}_{\alpha})$ exists, where
\begin{equation}
F_{k l} (e^{i \theta}, \mathcal{C}^{(n)}_{\alpha}) = \lim_{\rho
\uparrow 1} F_{k l} (\rho e^{i \theta}, \mathcal{C}^{(n)}_{\alpha})
\end{equation}

Also, by the properties of Hardy spaces, $F_{k l} (\,\cdot\,,
\mathcal{C}^{(n)}_{\alpha}) \in L^s (\partial \bbD)$ for any $s \in
(0,1)$. Since the distributions of $\alpha_0, \alpha_1, \ldots,
\alpha_{n-1}$ are rotationally invariant, we obtain that for any
fixed $e^{i \theta} \in \partial \bbD$, the radial limit $F_{k l}
(e^{i \theta}, \mathcal{C}^{(n)}_{\alpha})$ exists for almost every
$\alpha \in \Omega$.

The relation (\ref{b7}) gives
\begin{equation}
\sup_{\rho \in (0,1)} \bbE \left(\left|F_{k l} (\rho e^{i \theta},
\mathcal{C}^{(n)}_{\alpha})\right|^s \right) \leq C
\end{equation}

By taking $\rho \uparrow 1$ and using Fatou's lemma we get:
\begin{equation}
\bbE\left(\left| F_{k l} (e^{i \theta}, \mathcal{C}^{(n)}_{\alpha})
\right|^s\right) \leq C
\end{equation}
\end{proof}

Note that the argument from Lemma \ref{l1} works in the same way
when we replace the unitary matrix $\mathcal{C}^{(n)}_{\alpha}$ with
the unitary operator $\mathcal{C}_{\alpha}$ (corresponding to random
Verblunsky coefficients uniformly distributed in $D(0,r)$), so we
also have
\begin{equation} \lb{3.15.18}
\bbE\left(\left| F_{k l} (e^{i \theta}, \mathcal{C}_{\alpha})
\right|^s\right) \leq C
\end{equation} for any nonnegative integers $k, l$ and for any $e^{i
\theta} \in \partial \bbD$.

The next step is to prove that the expectations of the fractional
moments of the resolvent of $\mathcal{C}^{(n)}$ tend to zero on the
rows. We will start with the following lemma suggested to us by
Aizenman \cite{AizPC}:

\begin{lemma} \lb{l2} Let $\{X_n = X_n(\omega)\}_{n \geq 0}$, $\omega \in \Omega$ be a family of
positive random variables such that there exists a constant $C
> 0$ such that $\bbE (X_n) < C$ and, for almost any $\omega \in \Omega$, $\lim_{n
\to \infty} X_n(\omega) = 0$. Then, for any $s \in (0,1)$,
\begin{equation} \lb{b10}
\lim_{n \to \infty} \bbE(X_n^s) = 0
\end{equation}
\end{lemma}

\begin{proof}
Let $\varepsilon > 0$ and let $M > 0$ such that $M^{s-1} <
\varepsilon$. Observe that if $X_n (\omega) > M$, then
$X_n^s(\omega) < M^{s-1} X_n(\omega)$. Therefore
\begin{equation}
X_n^s(\omega) \leq X_n^s(\omega) \, \chi_{\{\omega; X_n(\omega)
\leq M\}}(\omega) + M^{s - 1} X_n(\omega)
\end{equation}
Clearly, $\bbE(M^{s - 1} X_n) \leq \varepsilon C$ and, using
dominated convergence,
\begin{equation}
\bbE (X_n^s \,
\chi_{\{\omega; X_n(\omega) \leq M\}}) \rightarrow 0 \qquad
\textrm{as} \quad n \to \infty
\end{equation}

We immediately get that for any $\varepsilon > 0$ we have
\begin{equation}
\limsup_{n \to \infty} \bbE (X^s_n) \leq \bbE (X_n^s \,
\chi_{\{\omega; X_n(\omega) \leq M\}}) + \varepsilon C
\end{equation}
so we can conclude that (\ref{b10}) holds.
\end{proof}

We will use Lemma \ref{l2} to prove that for any fixed $j$,
$\bbE\left(\left| F_{j, j+k} (e^{i \theta}, \mathcal{C}_{\alpha})
\right|^s\right)$ and $\bbE\left(\left| F_{j, j+k} (e^{i \theta},
\mathcal{C}^{n}_{\alpha}) \right|^s\right)$ converge to $0$ as $k
\to \infty$. From now on, it will be more convenient to work with
the resolvent $G$ instead of the Carath\'eodory function $F$.

\begin{lemma} \lb{l3} Let $\mathcal{C} = \mathcal{C}_{\alpha}$ be the random CMV matrix associated
to a family of Verblunsky coefficients $\{\alpha_n\}_{n \geq 0}$
with $\alpha_n$ i.i.d. random variables uniformly distributed in a
disk $D(0,r)$, $0 < r < 1$. Let $s \in (0,1)$, $z \in \bbD \cup
\partial \bbD$, and $j$ a positive integer. Then we have
\begin{equation} \lb{b11}
\lim_{k \to \infty} \bbE \left(\left|G_{j, j+k}(z, \mathcal{C}
\right)\right|^{s}) = 0
\end{equation}
\end{lemma}

\begin{proof}
For any fixed $z \in \bbD$, the rows and columns of
$G(z,\mathcal{C})$ are $l^2$ at infinity, hence converge to 0. Let
$s' \in (s,1)$. Then we get (\ref{b11}) applying Lemma \ref{l2} to
the random variables $X_k = \left|G_{j, j+k}(z, \mathcal{C})
\right|^{s'}$ and using the power $\frac{s}{s'} < 1$.

We will now prove (\ref{b11}) for $z = e^{i \theta} \in \partial
\bbD$. In order to do this, we will have to apply the heavy
machinery of transfer matrices and Lyapunov exponents developed in
\cite{OPUC2}. Thus, the transfer matrices corresponding to the CMV
matrix are
\begin{equation}
T_n(z) = A(\alpha_n,z) \ldots A(\alpha_0,z)
\end{equation} where $A(\alpha,z) = (1 - |\alpha|^2)^{-1/2}\left(%
\begin{smallmatrix}
  z & -\overline{\alpha} \\
  -\alpha z & 1 \\
\end{smallmatrix}%
\right)$ and the Lyapunov exponent is
\begin{equation}
\gamma(z) = \lim_{n \to \infty} \frac{1}{n} \log \|
T_n(z,\{\alpha_n\})\|
\end{equation} (provided this limit exists).

Observe that the common distribution $d \mu_{\alpha}$ of the
Verblunsky coefficients $\alpha_n$ is rotationally invariant and
\begin{equation}
\int_{D(0,1)} - \log (1 - \omega) \, d \mu_{\alpha} (\omega) <
\infty
\end{equation}
and
\begin{equation}
\int_{D(0,1)} - \log |\omega| \, d \mu_{\alpha} (\omega) < \infty
\end{equation}

Let us denote by $d \nu_N$ the density of eigenvalues measure and
let $U^{d \nu_N}$ be the logarithmic potential of the measure $d
\nu_N$, defined by
\begin{equation}
U^{d \nu_N} (e^{i \theta}) = \int_{\partial \bbD} \log
\frac{1}{|e^{i \theta} - e^{i \tau}|} \, d\nu_N(e^{i \tau})
\end{equation}

By rotation invariance, we have $d \nu_N = \frac{d \theta}{2 \pi}$
and therefore $U^{d \nu_N}$ is identically zero. Using results from
\cite{OPUC2}, the Lyapunov exponent exists for every $z = e^{i
\theta} \in
\partial \bbD$ and the Thouless formula gives
\begin{equation}
\gamma(z) = - \tfrac{1}{2} \, \int_{D(0,1)} \log (1 - |\omega|^2) \,
d\mu_{\alpha} (\omega)
\end{equation}

By an immediate computation we get $\gamma(z) = \frac{r^2 + (1 -
r^2) \log(1 - r^2)}{2 r^2} > 0$.

The positivity of the Lyapunov exponent $\gamma(e^{i \theta})$
implies (using the Ruelle-Osceledec theorem; see \cite{OPUC2}) that
there exists a constant $\lambda \neq 1$ (defining a boundary
condition) for which
\begin{equation}
\lim_{n \to \infty} T_n(e^{i \theta}) \binom{1}{\lambda} = 0
\end{equation}

From here we immediately get (using the theory of subordinate
solutions developed in \cite{OPUC2}) that for any $j$ and almost
every $e^{i \theta} \in \partial \bbD$,
\begin{equation} \lb{3.26.18}
\lim_{k \to \infty} G_{j, j+k}(e^{i \theta}, \mathcal{C}) = 0
\end{equation}

We can use now (\ref{3.15.18}) and (\ref{3.26.18}) to verify the
hypothesis of Lemma \ref{l2} for the random variables
\begin{equation}
X_k = \left|G_{j, j+k}(e^{i \theta}, \mathcal{C})\right|^{s'}
\end{equation} where $s' \in (s,1)$. We therefore get
\begin{equation}
\lim_{k \to \infty} \bbE \left(\left|G_{j, j+k}(e^{i \theta},
\mathcal{C} \right)\right|^{s}) = 0
\end{equation}
\end{proof}

The next step is to get the same result for the finite volume case
(i.e., when we replace the matrix $\mathcal{C} =
\mathcal{C}_{\alpha}$ by the matrix $\mathcal{C}_{\alpha}^{(n)}$).

\begin{lemma} \lb{l4} For any fixed $j$, any $s \in (0,\frac{1}{2})$, and any $z \in \bbD \cup \partial \bbD$,
\begin{equation} \lb{b13}
\lim_{k \to \infty, \, k\leq n} \bbE \left(\left|G_{j, j+k}(z,
\mathcal{C}^{(n)}_{\alpha} \right)\right|^{s}) = 0
\end{equation}
\end{lemma}

\begin{proof} Let $\mathcal{C}$ be the CMV matrix corresponding to a family of
Verblunsky coefficients $\{ \alpha_n \}_{n \geq 0}$, with
$|\alpha_n| < r$ for any $n$. Since $\bbE \left(\left|G_{j, j+k}(z,
\mathcal{C}) \right|^{s} \right) \rightarrow 0$ and $\bbE
\left(\left|G_{j, j+k}(z, \mathcal{C}) \right|^{2s} \right)
\rightarrow 0$ as $k \to \infty$, we can take $k_{\varepsilon} \geq
0$ such that for any $k \geq k_{\varepsilon}$, $\bbE
\left(\left|G_{j, j+k}(z, \mathcal{C}) \right|^{s} \right) \leq
\varepsilon$ and $\bbE \left(\left|G_{j, j+k}(z, \mathcal{C})
\right|^{2s} \right) \leq \varepsilon$.


For $n \geq (k_{\varepsilon} + 2)$, let $\mathcal{C}^{(n)}$ be the
CMV matrix obtained with the same $\alpha_0, \alpha_1, \ldots,
\alpha_{n-2}, \alpha_{n}, \ldots$ and with $\alpha_{n-1} \in
\partial \bbD$. From now on we will use $G(z, \mathcal{C}) = (\mathcal{C} - z)^{-1}$
and $G(z, \mathcal{C}^{(n)}_{\alpha}) = (\mathcal{C}^{(n)}_{\alpha}
- z)^{-1}$. Then
\begin{equation} \lb{b12}
(\mathcal{C}^{(n)}_{\alpha} - z)^{-1} - (\mathcal{C} - z)^{-1} =
(\mathcal{C} - z)^{-1} (\mathcal{C} - \mathcal{C}^{(n)}_{\alpha})
(\mathcal{C}^{(n)}_{\alpha} - z)^{-1}
\end{equation}

Note that the matrix $(\mathcal{C} - \mathcal{C}^{(n)})$ has at most
eight nonzero terms, each of absolute value at most 2. These nonzero
terms are situated at positions $(m,m')$ and $|m - n| \leq 2$, $|m'
- n| \leq 2$. Then
\begin{align}
\qquad \bbE \left(|(\mathcal{C}^{(n)}_{\alpha} -
z)^{-1}_{j,j+k}|^s\right) &\leq \bbE \left(|(\mathcal{C} -
z)^{-1}_{j,j+k}|^s\right) \notag\\& \quad + 2^s\, \sum_{8\, {\rm
terms}} \bbE \left(|(\mathcal{C} - z)^{-1}_{j,m}|^s \,
|(\mathcal{C}^{(n)}_{\alpha} - z)^{-1}_{m',j+k}|^s\right)
\end{align}

Using Schwartz's inequality,
\begin{equation}
\begin{split}
&\bbE \left(|\mathcal{C} - z)^{-1}_{j,m}|^s \,
|(\mathcal{C}^{(n)}_{\alpha} - z)^{-1}_{m',j+k}|^s\right) \leq \\
& \qquad \bbE \left(|\mathcal{C} - z)^{-1}_{j,m}|^{2s} \right)^{1/2}
\bbE \left(|\mathcal{C}^{(n)}_{\alpha} - z)^{-1}_{m',j+k}|^{2s}
\right)^{1/2}
\end{split}
\end{equation}

We clearly have $m \geq k_{\varepsilon}$ and therefore $\bbE
\left(|\mathcal{C} - z)^{-1}_{j,m}|^{2 s} \right) \leq \varepsilon$.
Also, from Lemma \ref{l1}, there exists a constant $C$ depending
only on $s$ such that $\bbE \left(|\mathcal{C}^{(n)}_{\alpha} -
z)^{-1}_{m',j+k}|^{2 s} \right) \leq C$.

Therefore, for any $k \geq k_{\varepsilon}$, $\bbE
\left(|(\mathcal{C}^{(n)}_{\alpha} - z)^{-1}_{j,j+k}|^s\right) \leq
\varepsilon + {\varepsilon}^{1/2} C$.

Since $\varepsilon$ is arbitrary, we obtain (\ref{b13}).
\end{proof}

Note that Lemma \ref{l4} holds for any $s \in (0,\frac{1}{2})$. The
result can be improved using a standard method:

\begin{lemma} \lb{l5}
For any fixed $j$, any $s \in (0,1)$, and any $z \in \bbD$,
\begin{equation} \lb{b14}
\lim_{k \to \infty, \, k\leq n} \bbE \left(\left|G_{j, j+k}(z,
\mathcal{C}^{(n)}_{\alpha} \right)\right|^{s}) = 0
\end{equation}
\end{lemma}

\begin{proof}
Let $s \in [\frac{1}{2},1), t \in (s,1), r \in (0, \frac{1}{2})$.
Then using the H\"older inequality for $p = \frac{t-r}{t-s}$ and for
$q = \frac{t-r}{s-r}$, we get
\begin{align}
\bbE \left(|(\mathcal{C}^{(n)}_{\alpha} - z)^{-1}_{j,j+k}|^s\right)
&= \bbE \left(|(\mathcal{C}^{(n)}_{\alpha} -
z)^{-1}_{j,j+k}|^{\frac{r(t - s)}{t - r}} \
|(\mathcal{C}^{(n)}_{\alpha} - z)^{-1}_{j,j+k}|^{\frac{t(s-r)}{t -
r}}\right) \notag\\
&\leq \left(\bbE \left(|(\mathcal{C}^{(n)}_{\alpha} -
z)^{-1}_{j,j+k}|^{r} \right)\right)^{\frac{t-s}{t-r}} \ \left(\bbE
\left(|(\mathcal{C}^{(n)}_{\alpha} -
z)^{-1}_{j,j+k}|^{t}\right)\right)^{\frac{s-r}{t-r}}
\end{align}

From Lemma \ref{l1}, $\bbE (|(\mathcal{C}^{(n)}_{\alpha} -
z)^{-1}_{j,j+k}|^{t})$ is bounded by a constant depending only on
$t$ and from Lemma \ref{l4}, $\bbE (|(\mathcal{C}^{(n)}_{\alpha} -
z)^{-1}_{j,j+k}|^{r} )$ tends to 0 as $k \to \infty$. We immediately
get (\ref{b14}).
\end{proof}

We can improve the previous lemma to get that the convergence to 0
of $\bbE (|(\mathcal{C}^{(n)}_{\alpha} - z)^{-1}_{j,j+k}|^s)$ is
uniform in row $j$.

\begin{lemma} \lb{unifdec}
For any $\varepsilon > 0$, there exists a $k_{\varepsilon} \geq 0$
such that, for any $s, k, j, n$, $s \in (0,1), k > k_{\varepsilon},
n
> 0, 0 \leq j \leq (n-1)$, and for any $z \in \bbD \cup
\partial \bbD$, we have
\begin{equation} \lb{eq3.25}
\bbE \left(\left|G_{j, j+k}\left(z, \mathcal{C}^{(n)}_{\alpha}
\right)\right|^{s}\right) < \varepsilon
\end{equation}
\end{lemma}

\begin{proof} As in the previous lemma, it is enough to prove the result for all $z \in
\bbD$. Suppose the matrix $\mathcal{C}^{(n)}$ is obtained from the
Verblunsky coefficients $\alpha_0, \alpha_1, \ldots, \alpha_{n-1}$.
Let's consider the matrix $\mathcal{C}^{(n)}_{{\rm dec}}$ obtained
from the same Verblunsky coefficients with the additional
restriction $\alpha_m = e^{i \theta}$ where $m$ is chosen to be
bigger but close to $j$ (for example $m = j + 3$). We will now
compare $(\mathcal{C}^{(n)} - z)^{-1}_{j,j+k}$ and
$(\mathcal{C}^{(n)}_{\rm dec} - z)^{-1}_{j,j+k}$. By the resolvent
identity,
\begin{align}
\left|(\mathcal{C}^{(n)} - z)^{-1}_{j,j+k} \right| &=
\left|(\mathcal{C}^{(n)} - z)^{-1}_{j,j+k} - (\mathcal{C}^{(n)}_{\rm
dec} - z)^{-1}_{j,j+k} \right| \\&\leq 2 \sum_{|l - m| \leq 2, |l' -
m| \leq 2} \left|(\mathcal{C}^{(n)} - z)^{-1}_{j,l}\right|\,\left|
(\mathcal{C}^{(n)}_{\rm dec} - z)^{-1}_{l',j+k}\right|
\end{align}

The matrix $(\mathcal{C}^{(n)}_{\rm dec} - z)^{-1}$ decouples
between $m-1$ and $m$. Also, since $|l' - m| \leq 2$, we get that
for any fixed $\varepsilon > 0$, we can pick a $k_{\varepsilon}$
such that for any $k \geq k_{\varepsilon}$ and any $l', |l' - m|
\leq 2$, we have
\begin{equation}
\bbE \left( \left| (\mathcal{C}^{(n)}_{\rm dec} - z)^{-1}_{l',j+k}
\right| \right) \leq \varepsilon
\end{equation}
(In other words, the decay is uniform on the 5 rows $m-2, m-1, m,
m+1$, and $m+2$ situated at distance at most 2 from the place where
the matrix $\mathcal{C}^{(n)}_{\rm dec}$ decouples.)

As in Lemma \ref{l4}, we can now use Schwartz's inequality to get
that for any $\varepsilon > 0$ and for any $s \in (0,\tfrac{1}{2})$
there exists a $k_{\varepsilon}$ such that for any $j$ and any $k
\geq k_{\varepsilon}$,
\begin{equation}
\bbE \left(\left| (\mathcal{C}^{(n)} - z)^{-1}_{j,j+k} \right|^s
\right) < \varepsilon
\end{equation}

Using the same method as in Lemma \ref{l5}, we get (\ref{eq3.25})
for any $s \in (0,1)$.

\end{proof}

We are heading towards proving the exponential decay of the
fractional moments of the matrix elements of the resolvent of the
CMV matrix. We will first prove a lemma about the behavior of the
entries in the resolvent of the CMV matrix.

\begin{lemma} \lb{l3.7} Suppose the random CMV matrix $\mathcal{C}^{(n)} = \mathcal{C}^{(n)}_{\alpha}$ is
given as before {\rm (}i.e., $\alpha_0, \alpha_1, \ldots,
\alpha_{n-2}, \alpha_{n-1}$ are independent random variables, the
first $(n-1)$ uniformly distributed inside a disk of radius $r$ and
the last one uniformly distributed on the unit circle{\rm)}. Then,
for any point $e^{i \theta} \in
\partial \bbD$ and for any $\alpha \in \Omega$ where $G(e^{i
\theta}, \mathcal{C}^{(n)}_{\alpha}) = (\mathcal{C}^{(n)}_{\alpha} -
e^{i \theta})^{-1}$ exists, we have
\begin{equation} \lb{concrho}
\frac{\left|G_{k l}(e^{i \theta},
\mathcal{C}^{(n)}_{\alpha})\right|}{\left|G_{i j}(e^{i \theta},
\mathcal{C}^{(n)}_{\alpha})\right|} \leq \left( \frac{2}{\sqrt{1 -
r^2}} \right)^{|k-i|+|l-j|}
\end{equation}
\end{lemma}

\begin{proof}
Using the results from Chapter 4 in Simon \cite{OPUC1}, the matrix
elements of the resolvent of the CMV matrix are given by the
following formulae:
\begin{equation} \lb{resfor}
\left[(\mathcal{C} - z)^{-1} \right]_{k l} = \left\{
\begin{array}{ll}
    (2z)^{-1} \chi_l(z) p_k(z), & k > l \quad {\rm or} \quad k = l = 2n - 1 \\
    (2z)^{-1} \pi_l(z) x_k(z), & l > k \quad {\rm or} \quad k = l = 2n \\
\end{array}
\right.
\end{equation} where the polynomials $\chi_l(z)$ are obtained by
the Gram-Schmidt process applied to $\{1, z, z^{-1}, \ldots \}$ in
$L^2(\partial \bbD, d \mu)$ and the polynomials $x_k(z)$ are
obtained by the Gram-Schmidt process applied to $\{1, z^{-1},z
\ldots \}$ in $L^2(\partial \bbD, d \mu)$. Also, $p_n$ and $\pi_n$
are the analogs of the Weyl solutions of Golinskii-Nevai \cite{GN}
and are defined by
\begin{align}
&p_n = y_n + F(z) x_n\\ &\pi_n = \Upsilon_n + F(z) \chi_n
\end{align}
where $y_n$ and $\Upsilon_n$ are the second kind analogs of the CMV
bases and are given by
\begin{equation}
y_n = \left\{%
\begin{array}{ll}
    z^{-l} \psi_{2l} & n=2l \\
    -z^{-l} \psi_{2l-1}^{*} & n=2l-1 \\
\end{array}%
\right.
\end{equation}
\begin{equation}
\Upsilon_n = \left\{%
\begin{array}{ll}
    -z^{-l} \psi_{2l}^{*} & n=2l \\
    z^{-l+1} \psi_{2l-1} & n=2l-1 \\
\end{array}%
\right.
\end{equation}

The functions $\psi_{n}$ are the second kind polynomials associated
to the measure $\mu$ and $F(z)$ is the Carath\'eodory function
corresponding to $\mu$ (see \cite{OPUC1}).

We will be interested in the values of the resolvent on the unit
circle (we know they exist a.e. for the random matrices considered
here). For any $z \in \partial \bbD$, the values of $F(z)$ are
purely imaginary and also $\overline{\chi_n(z)} = x_n(z)$ and
$\overline{\Upsilon_n(z)} = -y_n(z)$. In particular,
$|\overline{\chi_n(z)}| = |x_n(z)|$ for any $z \in \partial \bbD$.

Therefore $\overline{\pi_n(z)} = \overline{\Upsilon_n(z)} +
\overline{F(z) \chi_n(z)} = - p_n(z)$, so $|\overline{\pi_n(z)}| =
|p_n(z)|$ for any $z \in \partial \bbD$. We will also use $|\chi_{2
n + 1}(z)| = |\varphi_{2 n + 1} (z)|, |\chi_{2 n}(z)| = |\varphi_{2
n}^{*} (z)|, |x_{2 n}(z)| = |\varphi_{2 n} (z)|$, and $|x_{2 n -
1}(z)| = |\varphi_{2 n - 1}^{*}(z)|$ for any $z \in
\partial \bbD$. Also, from Section 1.5 in \cite{OPUC1}, we have
\begin{equation} \lb{keyfact}
\left| \frac{\varphi_{n \pm 1}(z)}{\varphi_{n}(z)} \right| \leq C
\end{equation} for any $z \in \partial \bbD$, where $C = 2/\sqrt{1 -
r^2}$.

The key fact for proving (\ref{keyfact}) is that the orthogonal
polynomials $\varphi_n$ satisfy a recurrence relation
\begin{equation} \lb{recforphi}
\varphi_{n+1}(z) = \rho_n^{-1} (z \varphi_n(z) - \overline{\alpha}_n
\varphi_n^{*}(z))
\end{equation}

This immediately gives the corresponding recurrence relation for the
second order polynomials
\begin{equation} \lb{recforpsi}
\psi_{n+1}(z) = \rho_n^{-1} (z \psi_n(z) + \overline{\alpha}_n
\psi_n^{*}(z))
\end{equation}

Using (\ref{recforphi}) and (\ref{recforpsi}), we will now prove a
similar recurrence relation for the polynomials $\pi_n$. For any $z
\in \partial \bbD$, we have
\begin{align} \lb{3.49.d}
\pi_{2l+1}(z) &= \Upsilon_{2l+1}(z) + F(z) \chi_{2l+1}(z) \notag\\
&= z^{-l} (\psi_{2l+1}(z) + F(z) \varphi_{2l+1}(z)) \notag\\
&= - \rho_{2l}^{-1} \, z \, \overline{\pi_{2l}(z)} + \rho_{2l}^{-1}
\, \overline{\alpha}_{2l} \, \pi_{2l}(z)
\end{align}
and similarly we get
\begin{align} \lb{3.50.d}
\pi_{2l}(z) = - \rho_{2l-1}^{-1} \overline{\pi_{2l-1}(z)} -
\alpha_{2l-1} \rho_{2l-1}^{-1} \pi_{2l-1}(z)
\end{align}
where we used the fact that for any $z \in \bbD$, $F(z)$ is purely
imaginary, hence $\overline{F(z)} = - F(z)$.

Since $\rho_{n}^{-1} \leq \frac{1}{\sqrt{1 - r^2}}$, the equations
(\ref{3.49.d}) and (\ref{3.50.d}) will give that for any integer $n$
and any $z \in \bbD$,
\begin{equation}
\left| \frac{\pi_{n \pm 1}(z)}{\pi_{n}(z)} \right| \leq C
\end{equation}
where $C = 2/ \sqrt{1 - r^2}$.

Using these observations and (\ref{resfor}) we get, for any $z \in
\partial \bbD$,
\begin{equation} \lb{Cright}
\left|\left[(\mathcal{C}^{(n)} - z)^{-1}\right]_{k,l}\right| \leq C
\left|\left[(\mathcal{C}^{(n)} - z)^{-1}\right]_{k,l \pm 1}\right|
\end{equation}
and also
\begin{equation} \lb{Cleft}
\left|\left[(\mathcal{C}^{(n)} - z)^{-1}\right]_{k,l}\right| \leq C
\left|\left[(\mathcal{C}^{(n)} - z)^{-1}\right]_{k \pm 1,l}\right|
\end{equation}

We can now combine (\ref{Cright}) and (\ref{Cleft}) to get
(\ref{concrho}).
\end{proof}

We will now prove a simple lemma which will be useful in
computations.

\begin{lemma} \lb{extralemma}
For any $s \in (0,1)$ and any constant $\beta \in \bbC$, we have
\begin{equation} \lb{extralemmaeq}
\int_{-1}^{1} \frac{1}{|x - \beta|^s} \, dx \leq \int_{-1}^{1}
\frac{1}{|x|^s} \, dx
\end{equation}
\end{lemma}

\begin{proof}
Let $\beta = \beta_1 + i \beta_2$ with $\beta_1, \beta_2 \in \bbR$.
Then
\begin{equation}
\int_{-1}^{1} \frac{1}{|x - \beta|^s} \, dx = \int_{-1}^{1}
\frac{1}{|(x - \beta_1)^2 + \beta_2^2|^{s/2}} \, dx \leq
\int_{-1}^{1} \frac{1}{|x - \beta_1|^s} \, dx
\end{equation}

But $1 / |x|^s$ is the symmetric decreasing rearrangement of $1 / |x
- \beta_1|^s$ so we get
\begin{equation}
\int_{-1}^{1} \frac{1}{|x - \beta_1|^s} \, dx \leq \int_{-1}^{1}
\frac{1}{|x|^s} \, dx
\end{equation} and therefore we immediately obtain
(\ref{extralemmaeq}).
\end{proof}

The following lemma shows that we can control conditional
expectations of the diagonal elements of the matrix
$\mathcal{C}^{(n)}$.

\begin{lemma} \lb{lemma3.9} For any $s
\in (0,1)$, any $k$, $1 \leq k \leq n$, and any choice of $\alpha_0,
\alpha_1, \ldots, \alpha_{k-1}, \alpha_{k+1}, \ldots, \alpha_{n-2},
\alpha_{n-1}$,
\begin{equation} \lb{3.20}
\bbE\left(\left| F_{k k} (z, \mathcal{C}^{(n)}_{\alpha}) \right|^s \
\big{|} \  \{ \alpha_i \}_{i \neq k} \right) \leq C
\end{equation} where a possible value for the constant is \, $C=\frac{4}{1-s} \,
32^s$.
\end{lemma}

\begin{proof}

For a fixed family of Verblunsky coefficients $\{ \alpha_n \}_{n
\in \bbN}$, the diagonal elements of the resolvent of the CMV
matrix $\mathcal{C}$ can be obtained using the formula:
\begin{equation}
(\delta_k, (\mathcal{C} + z)(\mathcal{C} - z)^{-1} \delta_k) =
\int_{\partial \bbD} \frac{e^{i \theta} + z}{e^{i \theta} -z} \,
|\varphi_k(e^{i \theta})|^2 \, d \mu (e^{i \theta})
\end{equation}
where $\mu$ is the measure on $\partial \bbD$ associated with the
Verblunsky coefficients $\{ \alpha_n \}_{n \in \bbN}$ and
$\{\varphi_n\}_{n \in \bbN}$ are the corresponding normalized
orthogonal polynomials.

\medskip

Using the results of Khrushchev \cite{Khr},
the Schur function of the measure $|\varphi_k(e^{i \theta})|^2 \,
d \mu (e^{i \theta})$ is:
\begin{equation}
g_k(z) = f(z; \alpha_k,\alpha_{k+1}, \ldots ) \, f(z;
-\overline{\alpha}_{k-1}, -\overline{\alpha}_{k-2}, \ldots,
-\overline{\alpha}_0,1)
\end{equation}
where by $f(z; S)$ we denote the Schur function associated to the
family of Verblunsky coefficients $S$.

\medskip

Since the dependence of $f(z; \alpha_k,\alpha_{k+1}, \ldots )$ \, on
$\alpha_k$ is given by
\begin{equation}
f(z; \alpha_k,\alpha_{k+1}, \ldots ) = \frac{\alpha_k + z f (z;
\alpha_{k+1}, \alpha_{k+2} \ldots)}{1 + \overline{\alpha}_k z f(z;
\alpha_{k+1}, \alpha_{k+2} \ldots)}
\end{equation}
we get that the dependence of $g_k(z)$ on $\alpha_k$ is given by
\begin{equation}
g_k(z) = C_1 \frac{\alpha_k \, + C_2}{1 + \overline{\alpha}_k C_2}
\end{equation}
where
\begin{align}
C_1 &= f(z; -\overline{\alpha}_{k-1}, -\overline{\alpha}_{k-2},
\ldots ,-\overline{\alpha}_0,1)\\ C_2 &= z f(z; \alpha_{k+1},
\alpha_{k+2}, \ldots)
\end{align}

\medskip

Note that the numbers $C_1$ and $C_2$ do not depend on $\alpha_k$,
$|C_1|, |C_2| \leq 1$.

\medskip

We now evaluate the Carath\'eodory function $F(z;|\varphi_k(e^{i
\theta})|^2 \, d \mu (e^{i \theta}))$ associated to the measure
$|\varphi_k(e^{i \theta})|^2 \, d \mu (e^{i \theta})$. By
definition,
\begin{align}
F(z;|\varphi_k(e^{i \theta})|^2 \, d \mu (e^{i \theta})) &=
\int_{\partial \bbD} \frac{e^{i \theta} + z}{e^{i \theta} -z} \,
|\varphi_k(e^{i \theta})|^2 \, d \mu (e^{i \theta}) \\&=
(\delta_k, (\mathcal{C} + z)(\mathcal{C} - z)^{-1} \delta_k)
\end{align}

We now have
\begin{align} \left|F(z;|\varphi_k(e^{i \theta})|^2 \, d \mu
(e^{i \theta}))\right| = \left|\frac{1 + z g_k(z)}{1 - z
g_k(z)}\right| \leq \left|\frac{2}{1 - z \, C_1 \, \frac{\alpha_k +
C_2}{1 + \overline{\alpha}_k C_2}}\right|
\end{align}

It suffices to prove
\begin{equation} \lb{3.32}
\sup_{w_1, w_2 \in \bbD} \int_{D(0,r)} \left|\frac{2}{1 - w_1 \,
\frac{\alpha_k + w_2}{1 + \overline{\alpha}_k w_2}}\right|^{s} \,
d \alpha_k \ < \  \infty
\end{equation}

Clearly
\begin{align} \lb{3.33}
\left|\frac{2}{1 - w_1 \, \frac{\alpha_k + w_2}{1 +
\overline{\alpha}_k w_2}}\right| &= \left|\frac{2 (1 +
\overline{\alpha}_k w_2)}{1 + \overline{\alpha}_k w_2 - w_1
(\alpha_k + w_2)} \right|\notag\\ &\leq \left|\frac{4}{1 +
\overline{\alpha}_k w_2 - w_1 (\alpha_k + w_2)} \right|
\end{align}

For $\alpha_k = x + i y$, $1 + \overline{\alpha}_k w_2 - w_1
(\alpha_k + w_2) = x (-w_1 + w_2) + y (- i w_1 - i w_2) + (1 - w_1
w_2)$. Since for $w_1, w_2 \in \bbD$, $(-w_1 + w_2)$, $(- i w_1 - i
w_2)$, and $(1 - w_1 w_2)$ cannot be all small, we will be able to
prove (\ref{3.32}).

If $| -w_1 + w_2 | \geq \varepsilon$,
\begin{align}
\int_{D(0,r)} \left|\frac{2}{1 - w_1 \, \frac{\alpha_k + w_2}{1 +
\overline{\alpha}_k w_2}}\right|^{s} \, d \alpha_k \leq
\left(\frac{4}{\varepsilon}\right)^{s} \int_{-r}^{r} \int_{-r}^{r}
\frac{1}{|x + y D + E|^s} \, dx \, dy\\
\leq 2 \left(\frac{4}{\varepsilon}\right)^{s} \int_{-1}^{1}
\frac{1}{|x|^s} \, dx = \frac{4}{1-s}
\left(\frac{4}{\varepsilon}\right)^{s}
\end{align}
(where for the last inequality we used Lemma \ref{extralemma}).

\smallskip

The same bound can be obtained for $| w_1 + w_2 | \geq
\varepsilon$.

If $| -w_1 + w_2 | \leq \varepsilon$ and $| w_1 + w_2 | \leq
\varepsilon$, then
\begin{equation}
|x (-w_1 + w_2) + y (- i w_1 - i w_2) + (1 - w_1 w_2)| \geq (1 -
\varepsilon^2 - 4 \varepsilon)
\end{equation}

so
\begin{equation}
\int_{D(0,r)} \left|\frac{2}{1 - w_1 \, \frac{\alpha_k + w_2}{1 +
\overline{\alpha}_k w_2}}\right|^{s} \, d \alpha_k \leq 2^{s+2}
\left(\frac{1}{1 - \varepsilon^2 - 4 \varepsilon}\right)^s
\end{equation}

Therefore for any small $\varepsilon$, we get $(\ref{3.20})$ with
\begin{equation}
C = \max \left\{ \frac{4}{1-s}
\left(\frac{4}{\varepsilon}\right)^{s}, 2^{s+2} \left(\frac{1}{1 -
\varepsilon^2 - 4 \varepsilon}\right)^s \right\}
\end{equation}

For example, for $\varepsilon = 1/8$, we get $C = \frac{4}{1-s} \,
32^s$.

\end{proof}

We will now be able to prove Theorem \ref{t1}.

\medskip

\begin{proof}[Proof of Theorem \ref{t1}] We will use the method developed by Aizenman et al.
\cite{Aiz_et_al} for Schr\"odinger operators. The basic idea is to
use the uniform decay of the expectations of the fractional moments
of the matrix elements of $\mathcal{C}^{(n)}$ (Lemma \ref{unifdec})
to derive the exponential decay.

We consider the matrix $\mathcal{C}^{(n)}$ obtained for the
Verblunsky coefficients $\alpha_0, \alpha_1, \ldots, \alpha_{n-1}$.
Fix a $k$, with $0 \leq k \leq (n-1)$. Let $\mathcal{C}^{(n)}_1$ be
the matrix obtained for the Verblunsky coefficients $\alpha_0,
\alpha_1, \ldots, \alpha_{n-1}$ with the additional condition
$\alpha_{k+m} = 1$ and $\mathcal{C}^{(n)}_2$ the matrix obtained
from $\alpha_0, \alpha_1, \ldots, \alpha_{n-1}$ with the additional
restriction $\alpha_{k+m+3} = e^{i \theta}$ ($m$ is an integer $\geq
3$ which will be specified later, and $e^{i \theta}$ is a random
point uniformly distributed on $\partial \bbD$).

Using the resolvent identity, we have
\begin{equation} \lb{res1}
(\mathcal{C}^{(n)} - z)^{-1} - (\mathcal{C}^{(n)}_1 - z)^{-1} =
(\mathcal{C}^{(n)}_1 - z)^{-1} \, (\mathcal{C}^{(n)}_1 -
\mathcal{C}^{(n)}) \, (\mathcal{C}^{(n)} - z)^{-1}
\end{equation} and
\begin{equation} \lb{res2}
(\mathcal{C}^{(n)} - z)^{-1} - (\mathcal{C}^{(n)}_2 - z)^{-1} =
(\mathcal{C}^{(n)} - z)^{-1} \, (\mathcal{C}^{(n)}_2 -
\mathcal{C}^{(n)}) \, (\mathcal{C}^{(n)}_2 - z)^{-1}
\end{equation}

Combining (\ref{res1}) and (\ref{res2}), we get
\begin{align}
(\mathcal{C}^{(n)} - z)^{-1} = (\mathcal{C}^{(n)}_1 - z)^{-1} +
(\mathcal{C}^{(n)}_1 - z)^{-1} \, (\mathcal{C}^{(n)}_1 -
\mathcal{C}^{(n)}) \, (\mathcal{C}^{(n)}_2 - z)^{-1} \notag \\
+ \, (\mathcal{C}^{(n)}_1 - z)^{-1} \, (\mathcal{C}^{(n)}_1 -
\mathcal{C}^{(n)}) \, (\mathcal{C}^{(n)} - z)^{-1} \,
(\mathcal{C}^{(n)}_2 - \mathcal{C}^{(n)}) \, (\mathcal{C}^{(n)}_2 -
z)^{-1}
\end{align}

For any $k, l$ with $l \geq (k + m)$, we have
\begin{equation}
\left[(\mathcal{C}^{(n)}_1 - z)^{-1}\right]_{k l} = 0
\end{equation} and
\begin{equation}
\left[(\mathcal{C}^{(n)}_1 - z)^{-1} \, (\mathcal{C}^{(n)}_1 -
\mathcal{C}^{(n)}) \, (\mathcal{C}^{(n)}_2 - z)^{-1}\right]_{k l} =
0
\end{equation}

Therefore, since each of the matrices $(\mathcal{C}^{(n)}_1 -
\mathcal{C}^{(n)})$ and $(\mathcal{C}^{(n)}_2 - \mathcal{C})$ has at
most eight nonzero entries, we get that
\begin{align}
\begin{split}
&\left[(\mathcal{C}^{(n)} - z)^{-1}\right]_{kl} = \\
&\sum_{64{\rm\ terms}} (\mathcal{C}^{(n)}_1 - z)^{-1}_{k s_1} \,
(\mathcal{C}^{(n)}_1 - \mathcal{C}^{(n)})_{s_1 s_2} \,
(\mathcal{C}^{(n)} - z)^{-1}_{s_2 s_3} \, (\mathcal{C}^{(n)}_2 -
\mathcal{C}^{(n)})_{s_3 s_4} \, (\mathcal{C}^{(n)}_2 - z)^{-1}_{s_4
l}
\end{split}
\end{align} which gives
\begin{equation}
\begin{split}
&\bbE \left(\left|(\mathcal{C}^{(n)} - z)^{-1}_{kl}\right|^s\right)
\\& \quad \leq 4^s \sum_{64{\rm\ terms}} \bbE \left(\left|(\mathcal{C}^{(n)}_1
- z)^{-1}_{k s_1} \, (\mathcal{C}^{(n)} - z)^{-1}_{s_2 s_3} \,
(\mathcal{C}^{(n)}_2 - z)^{-1}_{s_4 l}\right|^s \right)
\end{split}
\end{equation}
where, since the matrix $\mathcal{C}^{(n)}_1$ decouples at $(k+m)$,
we have $|s_2 - (k + m)| \leq 2$ and, since the matrix
$\mathcal{C}^{(n)}_1$ decouples at $(k+m+3)$, we have $|s_3 - (k + m
+ 3)| \leq 2$.

By Lemma \ref{l3.7}, we have for any $e^{i \theta} \in \partial
\bbD$,
\begin{equation}
\frac{\left|(\mathcal{C}^{(n)} - e^{i \theta})^{-1}_{s_2
s_3}\right|}{\left|(\mathcal{C}^{(n)} - e^{i \theta})^{-1}_{k+m+1,
k+m+1}\right|} \leq \left( \frac{2}{\sqrt{1 - r^2}} \right)^7
\end{equation}

Observe that $(\mathcal{C}^{(n)}_1 - z)^{-1}_{k s_1}$ and
$(\mathcal{C}^{(n)}_2 - z)^{-1}_{s_4 l}$ do not depend on
$\alpha_{k+m+1}$, and therefore using Lemma \ref{lemma3.9}, we get
\begin{align}
\bbE \left(\left|(\mathcal{C}^{(n)}_1 - z)^{-1}_{k s_1} \,
(\mathcal{C}^{(n)} - z)^{-1}_{s_2 s_3} \, (\mathcal{C}^{(n)}_2 -
z)^{-1}_{s_4 l}\right|^s \ \big{|} \  \{ \alpha_i \}_{i \neq
(k+m+1)}\right) \notag \\ \leq \frac{4}{1-s} \, 32^s \, \left(
\frac{2}{\sqrt{1 - r^2}} \right)^7 \left|(\mathcal{C}^{(n)}_1 -
z)^{-1}_{k s_1}\right|^s \left|(\mathcal{C}^{(n)}_2 - z)^{-1}_{s_4
l}\right|^s
\end{align}

Since the random variables $(\mathcal{C}^{(n)}_1 - z)^{-1}_{k s_1}$
and $(\mathcal{C}^{(n)}_2 - z)^{-1}_{s_4 l}$ are independent (they
depend on different sets of Verblunsky coefficients), we get
\begin{align}
\begin{split}
&\bbE \left(\left|(\mathcal{C}^{(n)}_1 - z)^{-1}_{k s_1} \,
(\mathcal{C}^{(n)} - z)^{-1}_{s_2 s_3} \, (\mathcal{C}^{(n)}_2 -
z)^{-1}_{s_4 l}\right|^s \right) \\ &\quad \leq C(s,r) \, \bbE
\left(\left|(\mathcal{C}^{(n)}_1 - z)^{-1}_{k s_1}\right|^s\right)
\bbE\left(\left|(\mathcal{C}^{(n)}_1 - z)^{-1}_{s_4
l}\right|^s\right)
\end{split}
\end{align}
where $C(s,r) = \frac{4}{1-s} \, 32^s \, \left( \frac{2}{\sqrt{1 -
r^2}} \right)^7$.

The idea for obtaining exponential decay is to use the terms \,$\bbE
(|(\mathcal{C}^{(n)}_1 - z)^{-1}_{k s_1}|^s)$ to get smallness and
the terms $\bbE (|(\mathcal{C}^{(n)}_1 - z)^{-1}_{s_4 l}|^s)$ to
repeat the process. Thus, using the Lemma \ref{unifdec}, we get that
for any $\beta < 1$, there exists a fixed constant $m \geq 0$ such
that, for any $s_1$, $|s_1 - (k + m)| \leq 2$, we have
\begin{equation}
4^s \cdot 64 \cdot C(s,r) \cdot \bbE
\left(\left|(\mathcal{C}^{(n)}_1 - z)^{-1}_{k s_1}\right|^s\right) <
\beta
\end{equation}

We can now repeat the same procedure for each term $\bbE
(|(\mathcal{C}^{(n)}_1 - z)^{-1}_{s_4 l}|^s)$ and we gain one more
coefficient $\beta$. At each step, we move $(m+3)$ spots to the
right from $k$ to $l$. We can repeat this procedure $\left[\frac{l -
k}{m+3}\right]$ times and we get
\begin{equation}
\bbE \left(\left|(\mathcal{C}^{(n)} - z)^{-1}_{kl}\right|^s\right)
\leq C \beta^{(l - k)/(m+3)}
\end{equation} which immediately gives (\ref{expdec}).
\end{proof}

\bigskip

\section{The Localized Structure of the Eigenfunctions.} \lb{s4}

In this section we will study the eigenfunctions of the random CMV
matrices considered in (\ref{model}). We will prove that, with
probability 1, each eigenfunction of these matrices will be
exponentially localized about a certain point, called the center of
localization. We will follow ideas from del Rio et al.
\cite{delRio}.

Theorem \ref{t1} will give that, for any $z \in \partial\bbD$, any
integer $n$ and any $s \in (0,1)$,
\begin{equation} \lb{expdecay}
\bbE \left(\left| F_{k l} (z, \mathcal{C}^{(n)}_{\alpha}) \right|^s
\right) \leq C \, e^{- D |k - l|} \end{equation}

Aizenman's theorem for CMV matrices (see Simon \cite{Sta}) shows
that (\ref{expdecay}) implies that for some positive constants $C_0$
and $D_0$ depending on $s$, we have
\begin{equation} \lb{4.2}
\bbE \left(\sup_{j \in \bbZ}\left| \left(\delta_k,
(\mathcal{C}^{(n)}_{\alpha})^j \delta_l \right) \right| \right) \leq
C_0 \, e^{- D_0 |k - l|}
\end{equation}

This will allow us to conclude that the eigenfunctions of the CMV
matrix are exponentially localized. The first step will be:

\begin{lemma} \lb{l4.1} For almost every $\alpha \in \Omega$, there exists a
constant $D_{\alpha} > 0$ such that for any $n$, any $k,l$, with $1
\leq k,l \leq n$, we have
\begin{equation} \lb{e4.2.1}
\sup_{j \in \bbZ} |(\delta_k, (\mathcal{C}^{(n)}_{\alpha})^{j}
\delta_l)| \leq D_{\alpha} \, (1 + n)^6 \, e^{-D_0 |k - l|}
\end{equation}
\end{lemma}

\begin{proof} From (\ref{4.2}) we get that \begin{equation} \int_{\Omega}
\left(\sup_{j \in \bbZ} \left| \left( \delta_k,
(\mathcal{C}^{(n)}_{\alpha})^j \delta_l \right) \right| \right) d
P(\alpha) \, \leq \, C_0 \, e^{- D_0 |k - l|}
\end{equation}
and therefore there exists a constant $C_1 > 0$ such that
\begin{equation} \lb{cond4.5}
\sum_{n,k,l=1\ k,l \leq n}^{\infty} \int_{\Omega} \frac{1}{(1 + n)^2
(1 + k)^2 (1 + l)^2} \left(\sup_{j \in \bbZ}\left| \left(\delta_k,
(\mathcal{C}^{(n)}_{\alpha})^j \delta_l \right) \right| \right)
e^{D_0 |k - l|} \ d P(\alpha) \, \leq \, C_1
\end{equation}

It is clear that for any $k,l$, with $1 \leq k,l \leq n$, the
function
\begin{equation}
\alpha \longrightarrow \frac{1}{(1 + n)^2 (1 + k)^2 (1 + l)^2}
\left(\sup_{j \in \bbZ}\left| \left(\delta_k,
(\mathcal{C}^{(n)}_{\alpha})^j \delta_l \right) \right| \right)
e^{D_0 |k - l|}
\end{equation} is integrable.

Hence, for almost every $\alpha \in \Omega$, there exists a constant
$D_{\alpha} > 0$ such that for any $n, k, l$, with $1 \leq k,l \leq
n$,
\begin{equation} \lb{4.7}
\sup_{j \in \bbZ}\left|\left(\delta_k,
(\mathcal{C}^{(n)}_{\alpha})^j \delta_l \right) \right| \ \leq\
D_{\alpha} \, (1 + n)^6 \, e^{-D_0 |k - l|}
\end{equation}
\end{proof}

A useful version of the previous lemma is:

\begin{lemma} \lb{l4.2}For almost every $\alpha \in \Omega$, there exists a
constant $C_{\alpha} > 0$ such that for any $n$, any $k,l$, with $1
\leq k,l \leq n$, and $|k-l| \geq \frac{12}{D_0} \ln (n+1)$, we have
\begin{equation} \lb{l4.1.1}
\sup_{j \in \bbZ} |(\delta_k, (\mathcal{C}^{(n)}_{\alpha})^{j}
\delta_l)| \leq C_{\alpha} \, e^{- \frac{D_0}{2} |k - l|}
\end{equation}
\end{lemma}

\begin{proof} It is clear that for any $n, k, l$, with $1 \leq k,l \leq n$
and $|k-l| \geq \frac{12}{D_0} \ln (n+1)$,

\begin{equation}
\frac{1}{(1 + n^2)(1 + k^2)(1 + l^2)} \  e^{\frac{D_0}{2} |k - l|}
\geq 1
\end{equation}

In particular, for any $n,k,l$ with $|k-l| \geq \frac{12}{D_0} \ln
(n+1)$, the function

\begin{equation}
\Omega \ni \alpha \ \longrightarrow \ \left(\sup_{j \in
\bbZ}\left|\left(\delta_k, (\mathcal{C}^{(n)}_{\alpha})^j \delta_l
\right) \right| \right) e^{\frac{D_0}{2} |k - l|}
\end{equation} is integrable, so it is finite for almost every $\alpha$.

Hence for almost every $\alpha \in \Omega$, there exists a constant
$C_{\alpha} > 0$ such that for any $k, l$, $|k-l| \geq
\frac{12}{D_0} \ln (n+1)$,
\begin{equation} \lb{4.11}
\sup_{j \in \bbZ}\left|\left(\delta_k,
(\mathcal{C}^{(n)}_{\alpha})^j \delta_l \right) \right| \ \leq\
C_{\alpha} \, e^{-\frac{D_0}{2} |k - l|}
\end{equation}

\end{proof}

\medskip

\begin{proof}[Proof of Theorem \ref{t2}] Let us start with a CMV matrix
$\mathcal{C}^{(n)} = \mathcal{C}^{(n)}_{\alpha}$ corresponding to
the Verblunsky coefficients $\alpha_0, \alpha_1, \ldots,
\alpha_{n-2}, \alpha_{n-1}$. As mentioned before, the spectrum of
$\mathcal{C}^{(n)}$ is simple. Let $e^{i \theta_{\alpha}}$ be an
eigenvalue of the matrix $\mathcal{C}^{(n)}_{\alpha}$ and
$\varphi^{(n)}_{\alpha}$ a corresponding eigenfunction.

We see that, on the unit circle, the sequence of functions
\begin{equation}
f_M (e^{i \theta}) = \frac{1}{2 M + 1} \sum_{j = -M}^{M} e^{i j
(\theta - \theta_{\alpha})}
\end{equation}
is uniformly bounded (by 1) and converge pointwise (as $M \to
\infty$) to the characteristic function of the point $e^{i
\theta_{\alpha}}$. Let $P_{\{ e^{i \theta_{\alpha}}\}} = \chi_{\{
e^{i \theta_{\alpha}}\}} (\mathcal{C}^{(n)}_{\alpha})$.

\bigskip

By Lemma \ref{l4.2}, we have, for any $k,l$, with $|k - l| \geq
\frac{12}{D_0} \ln (n+1)$,
\begin{align}
\left|\left(\delta_k, f_M(\mathcal{C}^{(n)}_{\alpha}) \, \delta_l
\right) \right| &= \frac{1}{2 M + 1} \left| \sum_{j=-M}^{M}
\left(\delta_k, e^{- i j\,
\theta_{\alpha}}(\mathcal{C}^{(n)}_{\alpha})^j \, \delta_l \right)
\right|\\
&\leq \ \frac{1}{2 M + 1} \sum_{j=-M}^{M} \left| \left(\delta_k,
(\mathcal{C}^{(n)}_{\alpha})^j \, \delta_l \right) \right| \ \leq\
C_{\alpha} \, e^{-\frac{D_0}{2} |k - l|}
\end{align}
where for the last inequality we used (\ref{l4.1}).

By taking $M \to \infty$ in the previous inequality, we get
\begin{equation}
\left|\left(\delta_k, P_{\{ e^{i \theta_{\alpha}}\}} \, \delta_l
\right) \right| \leq\ C_{\alpha} \, e^{-\frac{D_0}{2} |k - l|}
\end{equation}
and therefore
\begin{equation} \lb{4.14}
\left| \varphi^{(n)}_{\alpha}(k) \, \varphi^{(n)}_{\alpha}(l)
\right| \leq\ C_{\alpha} \, e^{-\frac{D_0}{2} |k - l|}
\end{equation}

We can now pick as the center of localization the smallest integer
$m({\varphi_{\alpha}^{(n)}})$ such that
\begin{equation}
|\varphi^{(n)}_{\alpha}(m({\varphi_{\alpha}^{(n)}}))| = \max_{m}
|\varphi^{(n)}_{\alpha}(m)|
\end{equation}
We clearly have
$|\varphi^{(n)}_{\alpha}(m({\varphi_{\alpha}^{(n)}}))| \geq
\frac{1}{\sqrt{n + 1}}$.

Using the inequality ({\ref{4.14}}) with $k = m$ and $l =
m({\varphi_{\alpha}^{(n)}})$ we get, for any $m$ with $|m -
m({\varphi_{\alpha}^{(n)}})| \geq \frac{12}{D_0} \ln (n+1)$,
\begin{equation}
\left| \varphi^{(n)}_{\alpha}(m) \right| \leq\ C_{\alpha} \,
e^{-\frac{D_0}{2} |m-m({\varphi_{\alpha}^{(n)}})|} \, \sqrt{n + 1}
\end{equation}

Since for large $n$, $e^{-\frac{D_0}{2} |k - l|} \, \sqrt{n + 1}
\leq e^{-\frac{D_0}{3} |k - l|}$ for any $k,l$, $|k-l| \geq
\frac{12}{D_0} \ln (n+1)$, we get the desired conclusion (we can
take $D_2 = \frac{12}{D_0}$).
\end{proof}

For any eigenfunction $\varphi_{\alpha}^{(n)}$, the point
$m({\varphi_{\alpha}^{(n)}})$ is called its center of localization.
The eigenfunction is concentrated (has its large values) near the
point $m({\varphi_{\alpha}^{(n)}})$ and is tiny at sites that are
far from $m({\varphi_{\alpha}^{(n)}})$. This structure of the
eigenfunctions will allow us to prove a decoupling property of the
CMV matrix.

Note that we used Lemma \ref{l4.2} in the proof of Theorem \ref{t2}.
We can get a stronger result by using Lemma \ref{l4.1} (we replace
(\ref{4.11}) by (\ref{4.7})). Thus, for any $n$ and any $m \leq n$,
we have
\begin{equation} \lb{4.19}
\left| \varphi^{(n)}_{\alpha}(m) \right| \leq\ D_{\alpha} \, (1 +
n)^6 \, e^{-\frac{D_0}{2} \, |m - m({\varphi_{\alpha}^{(n)}})|} \,
\sqrt{n + 1}
\end{equation}
where $m({\varphi_{\alpha}^{(n)}})$ is the center of localization of
the eigenfunction $\varphi^{(n)}_{\alpha}$.

\section{Decoupling the Point Process} \lb{s5}

We will now show that the distribution of the eigenvalues of the
CMV matrix $\mathcal{C}^{(n)}$ can be approximated (as $n \to
\infty$) by the distribution of the eigenvalues of another matrix
CMV matrix $\tilde{\mathcal{C}}^{(n)}$, which decouples into the
direct sum of smaller matrices.

As explained in Section \ref{s1}, for the CMV matrix
$\mathcal{C}^{(n)}$ obtained with the Verblunsky coefficients
$\alpha = (\alpha_0, \alpha_1, \ldots, \alpha_{n-1}) \in \Omega$, we
consider $\tilde{\mathcal{C}}^{(n)}$ the CMV matrix obtained from
the same Verblunsky coefficients with the additional restrictions
$\alpha_{\left[ \frac{n}{\ln n}\right]} = e^{i \eta_1}$, $\alpha_{2
\left[ \frac{n}{\ln n} \right]} = e^{i \eta_2}$, \ldots,
\,$\alpha_{n - 1} = e^{i \eta_{[\ln n]}}$, where $e^{i \eta_1}, e^{i
\eta_2}, \ldots, e^{i \eta_{[\ln n]}}$ are independent random points
uniformly distributed on the unit circle. The matrix
$\tilde{\mathcal{C}}^{(n)}$ decouples into the direct sum of
approximately \,$[\ln n]$ unitary matrices
$\tilde{\mathcal{C}}^{(n)}_1$, $\tilde{\mathcal{C}}^{(n)}_2$,
\ldots, $\tilde{\mathcal{C}}^{(n)}_{\left[ \ln n \right]}$. Since we
are interested in the asymptotic distribution of the eigenvalues, it
will be enough to study the distribution (as $n \to \infty$) of the
eigenvalues of the matrices $\mathcal{C}^{(N)}$ of size $N =
\left[\ln n\right] \left[\frac{n}{\ln n}\right]$. Note that in this
situation the corresponding truncated matrix
$\tilde{\mathcal{C}}^{(N)}$ will decouple into the direct sum of
exactly $[\ln n]$ identical blocks of size $\left[\frac{n}{\ln
n}\right]$.

We will begin by comparing the matrices $\mathcal{C}^{(N)}$ and
$\tilde{\mathcal{C}}^{(N)}$.

\begin{lemma} \lb{l5.1} For $N = \left[\ln n\right]
\left[\frac{n}{\ln n}\right]$, the matrix $\mathcal{C}^{(N)} -
\tilde{\mathcal{C}}^{(N)}$ has at most $4 [\ln n]$ nonzero rows.
\end{lemma}

\begin{proof}
In our analysis, we will start counting the rows of the CMV matrix
with row 0. A simple inspection of the CMV matrix shows that for
even Verblunsky coefficients $\alpha_{2k}$, only the rows $2k$ and
$2k + 1$ depend on $\alpha_{2k}$. For odd Verblunsky coefficients
$\alpha_{2k+1}$, only the rows $2k, 2k+1,2k+2,2k+3$ depend on
$\alpha_{2k+1}$.

Since in order to obtain the matrix $\tilde{\mathcal{C}}^{(N)}$ from
$\mathcal{C}^{(N)}$ we modify $[\ln n]$ Verblunsky coefficients
$\alpha_{\left[ \frac{n}{\ln n}\right]}, \alpha_{2 \left[
\frac{n}{\ln n} \right]}, \ldots, \,\alpha_{[\ln n] \left[
\frac{n}{\ln n} \right]}$, we immediately see that at most $4 [\ln
n]$ rows of $\mathcal{C}^{(N)}$ are modified.

Therefore $\mathcal{C}^{(N)} - \tilde{\mathcal{C}}^{(N)}$ has at
most $4 [\ln n]$ nonzero rows (and, by the same argument, at most 4
columns around each point where the matrix
$\tilde{\mathcal{C}}^{(N)}$ decouples).
\end{proof}

Since we are interested in the points situated near the places
where the matrix $\tilde{\mathcal{C}}^{(N)}$ decouples, a useful
notation will be
\begin{equation} S_N (K) = S^{(1)}(K) \cup S^{(2)}(K) \cup \cdots
\cup S^{([\ln n])}(K)
\end{equation}
where $S^{(k)}(K)$ is a set of $K$ integers centered at $k \left[
\frac{n}{\ln n} \right]$ (e.g., for $K = 2p$, $S^{(k)}(K) = \left\{
k \left[ \frac{n}{\ln n} \right] - p + 1, k \left[ \frac{n}{\ln n}
\right] - p + 2, \ldots k \left[ \frac{n}{\ln n} \right]
+p\right\}$). Using this notation, we also have
\begin{equation}
S_N(1) = \left\{ \left[ \frac{n}{\ln n}\right], 2 \left[
\frac{n}{\ln n}\right], \ldots, [\ln n] \left[ \frac{n}{\ln
n}\right] \right\}
\end{equation}

Consider the intervals $I_{N,k}$, $1 \leq k \leq m$, of size
$\frac{1}{N}$ near the point $e^{i \alpha}$ on the unit circle (for
example $I_{N,k} = (e^{i (\alpha + \frac{a_k}{N})}, e^{i (\alpha +
\frac{b_k}{N})})$), where $a_1 < b_1 \leq a_2 < b_2 \leq \cdots \leq
a_m < b_m$. We will denote by $\mathcal{N}_N (I)$ the number of
eigenvalues of $\mathcal{C}^{(N)}$ situated in the interval $I$, and
by $\tilde{\mathcal{N}}_N (I)$ the number of eigenvalues of
$\tilde{\mathcal{C}}^{(N)}$ situated in $I$. We will prove that, for
large $N$, $\mathcal{N}_N (I_{N,k})$ can be approximated by
$\tilde{\mathcal{N}}_N (I_{N,k})$, that is, for any integers $k_1,
k_2, \ldots, k_m \geq 0$, we have, for $N \rightarrow \infty$,
\begin{align}
&\big| \ \bbP (\mathcal{N}_N (I_{N,1}) = k_1,\, \mathcal{N}_N
(I_{N,2})
= k_2, \ldots, \mathcal{N}_N (I_{N,m}) = k_m ) \notag\\
&- \bbP (\tilde{\mathcal{N}}_N (I_{N,1}) = k_1,\,
\tilde{\mathcal{N}}_N (I_{N,2}) = k_2, \ldots, \tilde{\mathcal{N}}_N
(I_{N,m}) = k_m ) \big| \longrightarrow 0
\end{align}




Since, by the results in Section \ref{s4}, the eigenfunctions of the
matrix $\mathcal{C}^{(N)}$ are exponentially localized (supported on
a set of size $2T [ \ln (n + 1) ]$, where, from now on, $T =
\frac{14}{D_0}$), some of them will have the center of localization
near $S_N(1)$ (the set of points where the matrix
$\tilde{\mathcal{C}}^{(N)}$ decouples) and others will have centers
of localization away from this set (i.e., because of exponential
localization, inside an interval $\left(k \left[ \frac{n}{\ln
n}\right],(k+1) \left[ \frac{n}{\ln n}\right]\right)$).

Roughly speaking, each eigenfunction of the second type will
produce an ``almost" eigenfunction for one of the blocks of the
decoupled matrix $\tilde{\mathcal{C}}^{(N)}$. These eigenfunctions
will allow us to compare $\mathcal{N}_N (I_{N,k})$ and
$\tilde{\mathcal{N}}_N (I_{N,k})$.

We see that any eigenfunction with the center of localization
outside the set $S_N(4 T [\ln n])$ will be tiny on the set $S_N(1)$.
Therefore, if we want to estimate the number of eigenfunctions that
are supported close to $S_N(1)$, it will be enough to analyze the
number $b_{N,\alpha}$, where $b_{N,\alpha}$ = number of
eigenfunctions of $\mathcal{C}^{(N)}_{\alpha}$ with the center of
localization inside $S_N(4 T [\ln n])$ (we will call these
eigenfunctions ``bad eigenfunctions"). We will now prove that the
number $b_{N,\alpha}$ is small compared to $N$.

A technical complication is generated by the fact that in the
exponential localization of eigenfunctions given by
(\ref{e4.2.1}), the constant $D_{\alpha}$ depends on $\alpha \in
\Omega$. We define
\begin{equation}
\mathcal{M}_{K} = \left\{\alpha \in \Omega,\  \sup_{j \in \bbZ}
|(\delta_k, (\mathcal{C}^{(N)})^{j} \delta_l)| \leq K \, (1 + N)^6
\, e^{-D_0 |k - l|}  \right\}
\end{equation}

Note that for any $K > 0$, the set $\mathcal{M}_{K} \subset \Omega$
is invariant under rotation. Also, we can immediately see that the
sets $\mathcal{M}_{K}$ grow with $K$ and
\begin{equation} \lb{5.6}
\lim_{K \to \infty} \bbP(\mathcal{M}_{K}) = 1
\end{equation}

We will be able to control the number of ``bad eigenfunctions" for
$\alpha \in \mathcal{M}_{K}$ using the following lemma:

\begin{lemma} \lb{l5.3}
For any $K > 0$ and any $\alpha \in \mathcal{M}_{K}$, there exists
a constant $C_K > 0$ such that
\begin{equation} \lb{5.7}
b_{N,\alpha} \leq C_K \left(\ln (1 + N)\right)^2
\end{equation}
\end{lemma}

\begin{proof} For any $K > 0$, any $\alpha \in \mathcal{M}_{K}$, and any
eigenfunction $\varphi^{N}_{\alpha}$ which is exponentially
localized about a point $m(\varphi^{N}_{\alpha})$, we have, using
(\ref{4.19}),
\begin{equation}
\left| \varphi^{(N)}_{\alpha}(m) \right| \leq\ K \,
e^{-\frac{D_0}{2} \, |m-m({\varphi_{\alpha}^{(N)}})|} \, (1 + N)^6
\, \sqrt{1 + N}
\end{equation}

Therefore for any $m$ such that $|m - m(\varphi^N_{\alpha})| \geq
\left[\frac{14}{D_0} \ln (1 + N)\right]$, we have
\begin{align}
\sum_{|m - m(\varphi^N_{\alpha})| \geq \left[\frac{14}{D_0} \ln (1 +
N)\right]} \left| \varphi^{(N)}_{\alpha}(m) \right|^2 &\leq 2
(1+N)^{-14} \, (1 + N)^{13} \
\sum_{k=0}^{\infty} K^2 \, e^{- D_0 k} \notag\\
&\leq (1+N)^{-1} K^{2} \frac{2 e^{D_0}}{e^{D_0} - 1}
\end{align}

Therefore, for any fixed $K$ and $s$, we can find an $N_0 =
N_0(k,s)$ such that for any $N \geq N_0$,
\begin{align}
\sum_{|m - m(\varphi^N_{\alpha})| \leq \left[\frac{14}{D_0} \ln (1 +
N)\right]} \left| \varphi^{(N)}_{\alpha}(m) \right|^2 \geq
\tfrac{1}{2} \lb{5.10}
\end{align}

We will consider eigenfunctions $\varphi^N_{\alpha}$ with the center
of localization in $S_N\left(4 T [\ln N]\right)$. For a fixed
$\alpha \in \mathcal{M}_{K}$, we denote the number of these
eigenfunctions by $b_{N,\alpha}$. We denote by $\{\psi_1, \psi_2,
\ldots, \psi_{b_{N,\alpha}} \}$ the set of these eigenfunctions.
Since the spectrum of $\mathcal{C}^{(N)}$ is simple, this is an
orthonormal set.

Therefore if we denote by $\mathrm{card} (A)$ the number of elements
of the set $A$, we get
\begin{align}
\begin{split}
\sum_{m \in S\left(4 T [\ln N] + \left[\frac{14}{D_0} \ln (1 +
N)\right]\right)} &\sum_{i=1}^{b_{N, \alpha}} \ |\psi_i(m)|^2\\
&\leq {\rm card} \left\{S\left(4 T [\ln N] + \left[\frac{14}{D_0}
\ln (1 + N)\right]\right)\right\} \notag\\
& \leq \left(4 T + \frac{14}{D_0}\right) \left(\ln (1 + N)\right)^2
\end{split}
\end{align}

Also, from (\ref{5.10}), for any $N \geq N_0(K,s)$,
\begin{equation}
\sum_{m \in S\left(4 T [\ln N] + \left[\frac{14}{D_0} \ln (1 + N)
\right]\right)} \sum_{i=1}^{b_{N, \alpha}} \ |\psi_i(m)|^2 \geq
\tfrac{1}{2} \ b_{N,\alpha}
\end{equation}

Therefore, for any $K > 0$ and any $\alpha \in \mathcal{M}_K$, we
have, for $N \geq N_0(K,s)$,
\begin{equation}
b_{N,\alpha} \leq 2 \left(4 T + \frac{14}{D_0}\right) \left( \ln (1
+ N) \right)^2
\end{equation}
and we can now conclude (\ref{5.7}).
\end{proof}

Lemma \ref{l5.3} shows that for any $K \geq 0$, the number of ``bad
eigenfunctions" corresponding to $\alpha \in \mathcal{M}_K$ is of
the order $(\ln N)^2$ (hence small compared to $N$).

Since the distributions for our Verblunsky coefficients are taken to
be rotationally invariant, the distribution of the eigenvalues is
rotationally invariant. Therefore, for any interval $I_N$ of size
$\frac{1}{N}$ on the unit circle, and for any fixed set
$\mathcal{M}_K \subset \Omega$, the expected number of ``bad
eigenfunctions" corresponding to eigenvalues in $I_N$ is of size
$\frac{(\ln N)^2}{N}$. We then get that the probability of the event
``there are bad eigenfunctions corresponding to eigenvalues in the
interval $I_N$" converges to 0. This fact will allow us to prove

\begin{lemma} \lb{l5.4}
For any $K > 0$, any disjoint intervals $I_{N,1}, I_{N,2}, \ldots,
I_{N,m}$ {\rm(}each one of size $\frac{1}{N}$ and situated near the
point $e^{i \alpha}${\rm)} and any positive integers $k_1, k_2,
\ldots, k_m$, we have
\begin{align} \lb{5.14}
&\big|\ \bbP (\{\mathcal{N}_N (I_{N,1}) = k_1,\, \mathcal{N}_N
(I_{N,2})
= k_2, \ldots, \mathcal{N}_N (I_{N,m}) = k_m \} \cap \mathcal{M}_K ) \notag\\
&- \bbP (\{\tilde{\mathcal{N}}_N (I_{N,1}) = k_1,\,
\tilde{\mathcal{N}}_N (I_{N,2}) = k_2,\, \ldots \,
\tilde{\mathcal{N}}_N (I_{N,m}) = k_m \} \cap \mathcal{M}_K ) \big|
\longrightarrow 0
\end{align}
as $N \to \infty$.
\end{lemma}

\begin{proof} We will work with $\alpha \in \mathcal{M}_K$. We first observe that any ``good eigenfunction"
(i.e., an eigenfunction with the center of localization outside
$S_N\left(4 T [\ln N]\right)$) is tiny on $S_N(1)$.

Indeed, from (\ref{4.19}), for any eigenfunction
$\varphi_{\alpha}^{(N)}$ with the center of localization
$m(\varphi_{\alpha}^{(N)})$ and for any $m$ with $|m -
m(\varphi_{\alpha}^{(N)})| \geq \frac{18}{D_0} [\ln (N + 1)]$,
\begin{equation}
|\varphi_{\alpha}^{(N)}(m)| \leq K e^{- \frac{D_0}{2} |m -
m(\varphi_{\alpha}^{(N)})|} (1 + N)^6 \sqrt{1 + N}
\end{equation}

In particular, if the center of localization of
$\varphi_{\alpha}^{(N)}$ is outside $S_N\left(4 T [\ln N]\right)$,
then for all $m \in S_N(1)$, we have
\begin{equation} \lb{5.16}
|\varphi_{\alpha}^{(N)}(m)| \leq K (1 + N)^{-2}
\end{equation}

We will use the fact that if $N$ is a normal matrix, $z_0 \in \bbC$,
$\varepsilon > 0$, and $\varphi$ is a unit vector with
\begin{equation}
\| \, (N - z_0) \varphi \,\| < \varepsilon
\end{equation}
then $N$ has an eigenvalue in $\{ z \ |\ |z - z_0| < \varepsilon
\}$.

For any ``good eigenfunction" $\varphi_{\alpha}^{(N)}$, we have
$\mathcal{C}^{(N)}_{\alpha} \varphi_{\alpha}^{(N)} = 0$ and
therefore, using Lemma \ref{l5.1},
\begin{equation}
\|\tilde{\mathcal{C}}^{(N)}_{\alpha} \varphi_{\alpha}^{(N)}\| \leq 2
K [\ln N] (1 + N)^{-2}
\end{equation}

Therefore, for any interval $I_N$ of size $\frac{1}{N}$, we have
\begin{equation}
\mathcal{N}_N(I_N) \leq \tilde{\mathcal{N}}_N(\tilde{I}_N)
\end{equation}
where $\tilde{I}_N$ is the interval $I_N$ augmented by $2 K [\ln N]
(1 + N)^{-2}$.

Since $2 K [\ln N] (1 + N)^{-2} = o(\frac{1}{N})$, we can now
conclude that
\begin{equation}
\bbP \big(\big(\mathcal{N}_N(I_N) \leq
\tilde{\mathcal{N}}_N(I_N)\big) \cap \mathcal{M}_K\big) \rightarrow
1 \quad {\rm as} \quad  n \to \infty
\end{equation}

We can use the same argument (starting from the eigenfunctions of
$\tilde{\mathcal{C}}^{(N)}_{\alpha}$, which are also exponentially
localized) to show that
\begin{equation}
\bbP\big(\big(\mathcal{N}_N(I_N) \geq
\tilde{\mathcal{N}}_N(I_N)\big) \cap \mathcal{M}_K\big) \rightarrow
1 \quad {\rm as} \quad  n \to \infty
\end{equation}
so we can now conclude that
\begin{equation}
\bbP\big(\big(\mathcal{N}_N(I_N) = \tilde{\mathcal{N}}_N(I_N)\big)
\cap \mathcal{M}_K\big) \rightarrow 1 \quad {\rm as} \quad  n \to
\infty
\end{equation}

Instead of one interval $I_N$, we can take $m$ intervals $I_{N,1},
I_{N,2} \ldots, I_{N,m}$ so we get (\ref{5.14}).
\end{proof}

\begin{proof}[Proof of Theorem \ref{t3}] Lemma \ref{l5.4} shows that for any $K
> 0$, the distribution of the eigenvalues of the matrix
$\mathcal{C}^{(N)}$ can be approximated by the distribution of the
eigenvalues of the matrix $\tilde{\mathcal{C}}^{(N)}$ when we
restrict to the set $\mathcal{M}_K \subset \Omega$. Since by
(\ref{5.6}) the sets $\mathcal{M}_{K}$ grow with $K$ and $\lim_{K
\to \infty} \bbP(\mathcal{M}_{K}) = 1$, we get the desired result.
\end{proof}

\bigskip

\section{Estimating the Probability of Having Two or More Eigenvalues in an Interval} \lb{s6}


The results from the previous section show that the local
distribution of the eigenvalues of the matrix $\mathcal{C}^{(N)}$
can be approximated by the direct sum of the local distribution of
$[\ln n]$ matrices of size $\left[\frac{n}{\ln n}\right]$,
$\mathcal{C}^{(N)}_1, \mathcal{C}^{(N)}_2, \ldots,
\mathcal{C}^{(N)}_{[\ln n]}$. These matrices are decoupled and
depend on independent sets of Verblunsky coefficients; hence they
are independent.

For a fixed point $e^{i \theta_0} \in \partial \bbD$, and an
interval $I_N = (e^{i (\theta_0 + \frac{2 \pi a}{N})}, e^{i
(\theta_0 + \frac{2 \pi b}{N})})$, we will now want to control the
probability of the event ``$\mathcal{C}^{(N)}$ has $k$ eigenvalues
in $I_N$." We will analyze the distribution of the eigenvalues of
the direct sum of the matrices $\mathcal{C}^{(N)}_1,
\mathcal{C}^{(N)}_2, \ldots, \mathcal{C}^{(N)}_{[\ln n]}$. We will
prove that, as $n \to \infty$, each of the decoupled matrices
$\mathcal{C}^{(N)}_k$ can contribute (up to a negligible error) with
at most one eigenvalue in the interval $I_N$.

For any nonnegative integer $m$, denote by $A(m,\mathcal{C},I)$ the
event
\begin{equation}
A(m,\mathcal{C},I) = ``\mathcal{C} \ {\rm has\  at\  least} \  m \  {\rm eigenvalues\  in\  the\  interval}\  I"
\end{equation}
and by $B(m, \mathcal{C}, I)$ the event
\begin{equation}
B(m, \mathcal{C}, I) = ``\mathcal{C}\  {\rm has\  exactly}\ m \ {\rm
eigenvalues \ in \ the\ interval}\ I"
\end{equation}

In order to simplify future notations, for any point $e^{i \theta}
\in \partial \bbD$, we also define the event $M(e^{i \theta})$ to be
\begin{equation}
M(e^{i \theta}) = ``e^{i \theta}\ {\rm is\ an\ eigenvalue\ of}\
\mathcal{C}^{(N)}"
\end{equation}

We can begin by observing that the eigenvalues of the matrix
$\mathcal{C}^{(N)}$ are the zeros of the $N$-th paraorthogonal
polynomial (see (\ref{l2.1.1})):
\begin{equation}
\Phi_{N}(z,d\mu,\beta) = z \Phi_{N-1}(z,d\mu) - \overline{\beta}\,
\Phi_{N-1}^{*} (z,d\mu)
\end{equation}

Therefore we can consider the complex function
\begin{equation} \lb{for6.0}
B_N (z)= \frac{\beta\,z\,\Phi_{N-1}(z)}{\Phi_{N-1}^{*}(z)}
\end{equation}
which has the property that $\Phi_N (e^{i \theta}) = 0$ if and only
if $B_N(e^{i \theta}) = 1$.

By writing the polynomials $\Phi_{N-1}$ and $\Phi_{N-1}^{*}$ as
products of their zeros, we can see that the function $B_N$ is a
Blaschke product.

Let $\eta_N : [0, 2 \pi) \rightarrow \bbR$ be a continuous function
such that
\begin{equation} \lb{eq6.6}
B_N(e^{i \theta}) = e^{i \, \eta_N(\theta)}
\end{equation}
(we will only be interested in the values of the function $\eta_N$
near a fixed point $e^{i \theta_0} \in \partial \bbD$). Note that
for any fixed $\theta \in \partial \bbD$, we have that $\eta
(\theta)$ is a random variable depending on $\alpha = (\alpha_0,
\alpha_1, \ldots, \alpha_{N-2}, \alpha_{N-1} = \beta) \in \Omega$.

We will now study the properties of the random variable $\eta_N
(\theta) = \eta_N (\theta, \alpha_0, \alpha_1, \ldots, \alpha_{N-2},
\beta)$. Thus
\begin{lemma} For any $\theta_1$ and $\theta_2$, the random variables
$\frac{\partial \eta_N}{\partial \theta}(\theta_1)$ and
$\eta_N(\theta_2)$ are independent. Also for any fixed value $w \in
\bbR$,
\begin{equation}
\bbE \left(\frac{\partial \eta_N}{\partial \theta}(\theta_1) \,
\Big| \, \eta_N (\theta_2) = w \right) = N
\end{equation}

\end{lemma}

\begin{proof}
The equation (\ref{for6.0}) gives
\begin{equation}
\eta_N(\theta) = \gamma + \tau(\theta)
\end{equation}
where $e^{i \gamma} = \beta$ and $e^{i \tau(\theta)} = \frac{e^{i
\theta}\,\Phi_{N-1}(e^{i \theta})}{\Phi_{N-1}^{*}(e^{i \theta})}$.
Since the distribution of each of the random variables $\alpha_0,
\alpha_1, \ldots, \alpha_{N-2}$ and $\beta$ is rotationally
invariant, for any $\theta \in [0, 2 \pi)$, $\gamma$ and
$\tau(\theta)$ are random variables uniformly distributed. Also, it
is immediate that $\gamma$ and $\tau(\theta)$ are independent. Since
$\gamma$ does not depend on $\theta$, for any fixed $\theta_1,
\theta_2 \in [0, 2 \pi)$, we have that the random variables
$\frac{\partial \eta_N}{\partial \theta}(\theta_1)$ and
$\eta_N(\theta_2)$ are independent.

We see now that for any Blaschke factor $B_a(z) = \frac{z - a}{1 -
\overline{a} z}$, we can define a real-valued function $\eta_a$ on
$\partial \bbD$ such that
\begin{equation}
e^{\eta_a (\theta)} = B_a(e^{i \theta})
\end{equation}

A straightforward computation gives
\begin{equation}
\frac{\partial \eta_a}{\partial \theta} (\theta) = \frac{1 -
|a|^2}{|e^{i \theta} - a|^2} > 0
\end{equation}

Since $B_N$ is a Blaschke product, we now get that for any fixed
$\alpha \in \Omega$, $\frac{\partial \eta_N}{\partial \theta}$ has a
constant sign (positive). This implies that the function $\eta_N$ is
strictly increasing. The function $B_N (z)$ is analytic and has
exactly $N$ zeros in $\bbD$ and therefore we get, using the argument
principle, that
\begin{equation}
\int_{0}^{2 \pi} \frac{\partial \eta_N}{\partial \theta} (\theta)\,
d \theta = 2 \pi N
\end{equation}

Note that $\frac{\partial \eta_N}{\partial \theta}$ does not depend
on $\beta$ (it depends only on $\alpha_0, \alpha_1, \ldots,
\alpha_{N-2}$). Also, using the same argument as in Lemma \ref{l1},
we have that for any angles $\theta$ and $\varphi$,
\begin{equation} \lb{eq6.11}
\frac{\partial \eta_N}{\partial \theta} (\theta) = \frac{\partial
\tilde{\eta}_N}{\partial \theta} (\theta - \varphi)
\end{equation}
where $\tilde{\eta}$ is the function $\eta$ that corresponds to the
Verblunsky coefficients
\begin{equation}
\alpha_{k, \varphi} = e^{- i (k+1) \varphi} \alpha_k \qquad k = 0,1,
\ldots, (N-2)
\end{equation}

Since the distribution of $\alpha_0, \alpha_1, \ldots, \alpha_{N-2}$
is rotationally invariant, we get from (\ref{eq6.11}) that the
function $\theta \rightarrow \bbE \left(\frac{\partial
\eta_N}{\partial \theta}(\theta)\right)$ is constant.

Taking expectations and using Fubini's theorem (as we also did in
Lemma \ref{l1}), we get, for any angle $\theta_0$,
\begin{equation}
2 \pi N = \bbE \left( \int_{0}^{2 \pi} \frac{\partial
\eta_N}{\partial \theta} (\theta)\, d \theta \right) = \int_{0}^{2
\pi} \bbE \left(\frac{\partial \eta_N}{\partial \theta}
(\theta)\right) \, d \theta = 2 \pi \, \bbE \left(\frac{\partial
\eta_N}{\partial \theta} (\theta_0)\right)
\end{equation}
and therefore
\begin{equation} \lb{6.81} \bbE \left(\frac{\partial
\eta_N}{\partial \theta}(\theta_0)\right) = N
\end{equation}

Since for any $\theta_1, \theta_2 \in [0, 2 \pi)$, we have that
$\frac{\partial \eta_N}{\partial \theta} (\theta_1)$ and $\eta_N
(\theta_2)$ are independent, (\ref{6.81}) implies that for any fixed
value $w \in \bbR$,
\begin{equation} \lb{6.91}
\bbE \left(\frac{\partial \eta_N}{\partial \theta}(\theta_1) \,
\Big| \, \eta_N (\theta_2) = w \right) = N
\end{equation}
\end{proof}

We will now control the probability of having at least two
eigenvalues in $I_N$ conditioned by the event that we already have
an eigenvalue at one fixed point $e^{i \theta_1} \in I_N$. This will
be shown in the following lemma:

\begin{lemma} \lb{l6.1} With $\mathcal{C}^{(N)}, I_N$, and the events $A(m,\mathcal{C},I)$ and
$M(e^{i \theta})$ defined before, and for any $e^{i \theta_1} \in
I_N$, we have
\begin{equation}
\bbP\left(A\left(2,\mathcal{C}^{(N)}, I_N) \ |\ M(e^{i
\theta_1}\right)\right) \leq (b-a)
\end{equation}
\end{lemma}

\begin{proof} Using the fact that the function $\theta \rightarrow \bbE \left(\frac{\partial \eta_N}{\partial
\theta}(\theta)\right)$ is constant and the relation (\ref{6.91}),
we get that
\begin{equation} \lb{6.12}
 \bbE \left(\int_{\theta_0 + \frac{2 \pi a}{N}}^{\theta_0 + \frac{2 \pi b}{N}} \frac{\partial \eta_N}{\partial
\theta}(\theta_1) \ d \theta_1 \, \Big| \, \eta_N (\theta_2) = w
\right)  = 2 \pi \,(b - a)
\end{equation}

We see that
\begin{equation}
\Phi_N(e^{i \theta}) = 0 \iff B_N (e^{i \theta}) = 1 \iff \eta_N
(\theta) = 0 \ ({\rm mod}\ 2 \pi)
\end{equation}

Therefore if the event $A(2,\mathcal{C}^{(N)}, I_N)$ takes place
(i.e., if the polynomial $\Phi_N$ vanishes at least twice in the
interval $I_N$), then the function $\eta_N$ changes by at least $2
\pi$ in the interval $I_N$, and therefore we have that
\begin{equation}
\int_{\theta_0 + \frac{2 \pi a}{N}}^{\theta_0 + \frac{2 \pi b}{N}}
\frac{\partial \eta_N}{\partial \theta}(\theta) \, d \theta \geq 2
\pi
\end{equation}
whenever the event $A(2,\mathcal{C}^{(N)}, I_N)$ takes place.

For any $\theta_1 \in I_N$ we have, using the independence of the
random variables $\frac{\partial \eta_N}{\partial \theta}(\theta_1)$
and $\eta_N(\theta_2)$ for the first inequality and Chebyshev's
inequality for the second inequality,
\bigskip
\begin{align} \lb{6.13}
\bbP\left(A(2,\mathcal{C}^{(N)}, I_N) \, \Big| \, M(e^{i \theta_1})
\right) &\leq \bbP\left(\int_{\theta_0 + \frac{2 \pi
a}{N}}^{\theta_0 + \frac{2 \pi b}{N}} \frac{\partial
\eta_N}{\partial \theta} (\theta) \, d \theta \geq 2 \pi
\, \Big| \, M(e^{i \theta_1}) \right) \notag\\
&\leq \frac{1}{2 \pi} \, \bbE \left(\int_{\theta_0 + \frac{2 \pi
a}{N}}^{\theta_0 + \frac{2 \pi b}{N}} \frac{\partial
\eta_N}{\partial \theta} (\theta) \, d \theta \, \Big| \, M(e^{i
\theta_1}) \right)
\end{align}

The previous formula shows that we can control the probability of
having more than two eigenvalues in the interval $I_N$ conditioned
by the event that a fixed $e^{i \theta_1}$ is an eigenvalue. We now
obtain, using $(\ref{6.12})$ with $w = 2 \pi m$, $m \in \bbZ$,
\begin{equation}
\bbP\left(A\left(2,\mathcal{C}^{(N)}, I_N) \ |\ M(e^{i
\theta_1}\right)\right) \leq (b-a)
\end{equation}
\end{proof}

We can now control the probability of having two or more eigenvalues in $I_N$.

\begin{theorem} \lb{t6.2} With $\mathcal{C}^{(N)}, I_N$, and the event $A(m,\mathcal{C},I)$ defined
before, we have
\begin{equation}
\bbP\left( A\left(2,\mathcal{C}^{(N)}, I_N \right)\right) \leq
\frac{(b-a)^2}{2}
\end{equation}
\end{theorem}

\begin{proof} For any positive integer $k$, we have
\begin{equation}
\bbP \left(B(k, \mathcal{C}^{(N)}, I_N)\right) = \frac{1}{k}
\int_{\theta_0 + \frac{2 \pi a}{N}}^{\theta_0 + \frac{2 \pi b}{N}}
\bbP \left(B(k, \mathcal{C}^{(N)}, I_N) \ \big| \ M(e^{i
\theta})\right) \, N \, d \nu_N (\theta)
\end{equation}
(where the measure $\nu_N$ is the density of eigenvalues).

Note that the factor $\frac{1}{k}$ appears because the selected
point $e^{i \theta}$ where we take the conditional probability can
be any one of the $k$ points.

\medskip

We will now use the fact that the distribution of the Verblunsky
coefficients is rotationally invariant and therefore for any $N$ we
have $d \nu_N = \frac{d \theta}{2 \pi}$, where $\frac{d \theta}{2
\pi}$ is the normalized Lebesgue measure on the unit circle.

Since for any $k \geq 2$ we have $\frac{1}{k} \leq \frac{1}{2}$, we
get that for any integer $k \geq 2$ and for large $N$,
\begin{equation}
\bbP \left(B(k, \mathcal{C}^{(N)}, I_N)\right) \leq \frac{N}{2}
\int_{\theta_0 + \frac{2 \pi a}{N}}^{\theta_0 + \frac{2 \pi b}{N}}
\bbP \left(B(k, \mathcal{C}^{(N)}, I_N) \ \big| \ M(e^{i
\theta})\right) \,\frac{d \theta}{2 \pi}
\end{equation}
and therefore,
\begin{equation}
\bbP \left(A(2, \mathcal{C}^{(N)}, I_N)\right) \leq \frac{N}{2}
\int_{\theta_0 + \frac{2 \pi a}{N}}^{\theta_0 + \frac{2 \pi b}{N}}
\bbP \left(A(2, \mathcal{C}^{(N)}, I_N) \ \big| \ M(e^{i
\theta})\right) \, \frac{d \theta}{2 \pi}
\end{equation}

Using Lemma \ref{l6.1}, we get
\begin{equation}
\bbP (A(2, \mathcal{C}^{(N)}, I_N)) \leq \frac{N}{2} \,
\frac{(b-a)}{N} \, (b-a) = \frac{(b-a)^2}{2}
\end{equation}

\end{proof}

\begin{theorem} \lb{t6.3} With $\mathcal{C}^{(N)}, \mathcal{C}^{(N)}_1,
\mathcal{C}^{(N)}_2, \ldots, \mathcal{C}^{(N)}_{[\ln n]}, I_N$, and the
event $A(m,\mathcal{C},I)$ defined
before, we have, for any $k$, $1 \leq k \leq [\ln n]$,
\begin{equation} \lb{c6.3}
\bbP\left(A(2,\mathcal{C}^{(N)}_k, I_N) \right) = O\left( \left([\ln
n]\right)^{-2} \right) \qquad {\rm as} \quad n \rightarrow \infty
\end{equation}
\end{theorem}

\begin{proof}
We will use the previous theorems for the CMV matrix
$\mathcal{C}^{(N)}_k$. Recall that $N = [\ln n] \left[\frac{n}{\ln
n}\right]$. Since this matrix has $\left[\frac{n}{\ln n}\right]$
eigenvalues, we can use the proof of Lemma \ref{l6.1} to obtain that
for any $e^{i \theta} \in I_N$,
\begin{equation}
\bbP\left(A(2,\mathcal{C}^{(N)}_k, I_N) \, \big| \, M(e^{i
\theta})\right) \leq \frac{1}{2 \pi} \, \frac{2 \pi (b-a)}{N} \,
 \left[\frac{n}{\ln n}\right] = \frac{b-a}{[\ln n]}
\end{equation}

The proof of Theorem \ref{t6.2} now gives
\begin{equation}
\bbP\left(A(2,\mathcal{C}^{(N)}_k, I_N) \right) \leq
\frac{(b-a)^2}{2 \, [\ln n]^2}
\end{equation}
and hence (\ref{c6.3}) follows.
\end{proof}

This theorem shows that as $N \to \infty$, any of the decoupled matrices contributes with at most one eigenvalue in each
interval of size $\frac{1}{N}$.

\section{Proof of the Main Theorem}

We will now use the results of Sections \ref{s3}, \ref{s4},
\ref{s5}, and \ref{s6} to conclude that the statistical distribution
of the zeros of the random paraorthogonal polynomials is Poisson.


\medskip

\begin{proof}[Proof of Theorem \ref{main}] It is enough to study the
statistical distribution of the zeros of polynomials of degree $N =
\left[ \ln n \right] \left[ \frac{n}{\ln n} \right]$. These zeros
are exactly the eigenvalues of the CMV matrix $\mathcal{C}^{(N)}$,
so, by the results in Section \ref{s5}, the distribution of these
zeros can be approximated by the distribution of the direct sum of
the eigenvalues of $\left[ \ln n \right]$ matrices
$\mathcal{C}^{(N)}_1, \mathcal{C}^{(N)}_2, \ldots,
\mathcal{C}^{(N)}_{[\ln n]}$.

In Section \ref{s6} (Theorem \ref{t6.3}), we showed that the
probability that any of the matrices $\mathcal{C}^{(N)}_1,
\mathcal{C}^{(N)}_2, \ldots, \mathcal{C}^{(N)}_{[\ln n]}$
contributes with two or more eigenvalues in each interval of size
$\frac{1}{N}$ situated near a fixed point $e^{i \theta} \in \partial
\bbD$ is of order $O (\left[ \ln n \right]^{-2})$. Since the
matrices $\mathcal{C}^{(N)}_1, \mathcal{C}^{(N)}_2, \ldots,
\mathcal{C}^{(N)}_{[\ln n]}$ are identically distributed and
independent, we immediately get that the probability that the direct
sum of these matrices has two or more eigenvalues in an interval of
size $\frac{1}{N}$ situated near $e^{i \theta}$ is $\left[\ln
n\right] O(\left[\ln n\right]^{-2})$ and therefore converges to 0 as
$n \to \infty$.

We can now conclude that as $n \rightarrow \infty$, the local
distribution of the eigenvalues converges to a Poisson process with
intensity measure $n \, \frac{d \theta}{2 \pi}$ using a standard
technique in probability theory. We first fix an interval $I_N =
(e^{i (\theta_0 + \frac{2 \pi a}{N})}, e^{i (\theta_0 + \frac{2 \pi
b}{N})})$ near the point $e^{i \theta_0}$ (as before, we take $N =
\left[\ln n\right] \left[ \frac{n}{\ln n} \right]$). Let us consider
$\left[\ln n \right]$ random variables $X_1, X_2, \ldots, X_{[\ln
n]}$ where $X_k$ = number of the eigenvalues of the matrix
$\mathcal{C}_k^{(N)}$ situated in the interval $I_N$ and let $S_n
(I_N) = X_1 + X_2 + \cdots + X_{[\ln n]}$. Note that $S_n (I_N)$ =
the number of eigenvalues of the matrix $\tilde{\mathcal{C}}^{(N)}$
situated in the interval $I_N$. We want to prove that
\begin{equation} \lb{rez7}
\lim_{n \to \infty} \bbP (S_n (I_N)= k) = e^{-(b-a)}
\frac{(b-a)^k}{k!}
\end{equation}

Theorem \ref{t6.3} shows that we can assume without loss of
generality that for any $k, 1 \leq k \leq [\ln n]$, we have $X_k \in
\{0,1\}$. Also, because of rotation invariance, we can assume, for
large $n$,
\begin{align}
\bbP( X_k = 1 ) &= \frac{(b - a)}{[\ln n]}\\
\bbP( X_k = 0 ) &= 1 - \frac{(b - a)}{[\ln n]}
\end{align}

The random variable $S_n (I_N)$ can now be viewed as the sum of
$[\ln n]$ Bernoulli trials, each with the probability of success
$\frac{(b - a)}{[\ln n]}$ and
\begin{equation}
\bbP (S_n (I_N)= k) = \binom{[\ln n]}{k} \left(\frac{(b - a)}{[\ln
n]}\right)^k \left(1 - \frac{(b - a)}{[\ln n]}\right)^{[\ln n] - k}
\end{equation}
which converges to $e^{-\lambda} \frac{\lambda^k}{k!}$, where
$\lambda = [\ln n] \frac{(b - a)}{[\ln n]} = ( b - a )$. Therefore
we get (\ref{rez7}).

Since for any disjoint intervals $I_{N,k}, 1 \leq k \leq [\ln n]$
situated near $e^{i \theta_0}$, the random variables $S_n (I_{N,k})$
are independent, (\ref{rez7}) will now give (\ref{CMT}) and
therefore the proof of the main theorem is complete.
\end{proof}

\section {Remarks}

1. We should emphasize the fact that the distribution of our
random Verblunsky coefficients is rotationally invariant. This
assumption is used in several places and seems vital for our
approach. It is not clear how (or whether) the approach presented
here can be extended to distributions that are not rotationally
invariant.

2. In this paper, we study the statistical distribution of the zeros
of paraorthogonal polynomials. It would be interesting to understand
the statistical distribution of the zeros of orthogonal polynomials.
A generic plot of the zeros of paraorthogonal polynomials versus the
zeros of orthogonal polynomials is
\begin{center}
\includegraphics[scale=.5,angle=0]{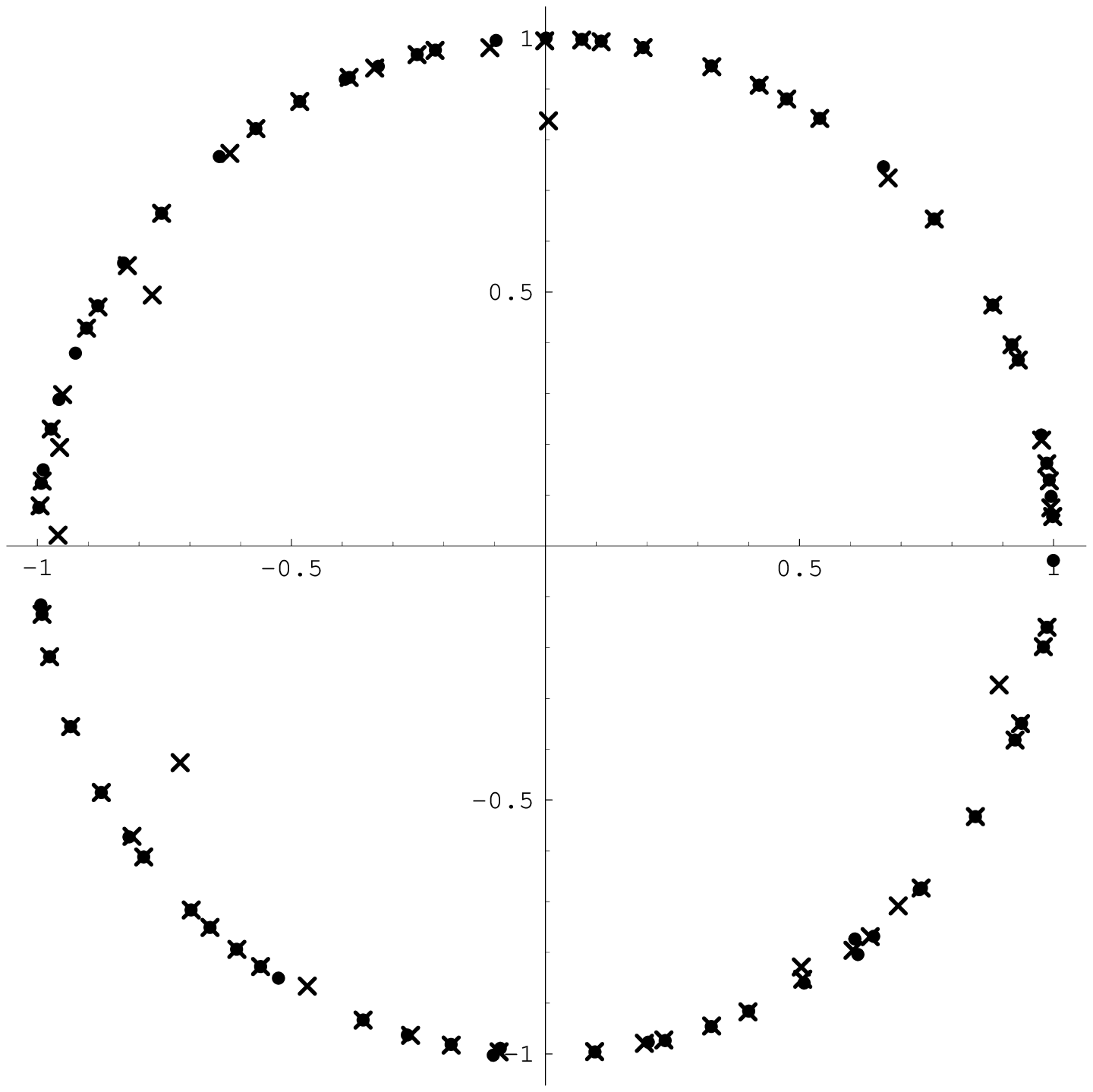}
\end{center}

In this Mathematica plot, the points represent the zeros of
paraorthogonal polynomials obtained by randomly choosing $\alpha_0,
\alpha_1, \ldots, \alpha_{69}$ from the uniform distribution on
$D(0,\frac{1}{2})$ and $\alpha_{70}$ from the uniform distribution
on $\partial \bbD$. The crosses represent the zeros of the
orthogonal polynomials obtained from the same $\alpha_0, \alpha_1,
\ldots, \alpha_{69}$ and an $\alpha_{70}$ randomly chosen from the
uniform distribution on $D(0, \frac{1}{2})$.

We observe that, with the exception of a few points (corresponding
probably to ``bad eigenfunctions"), the zeros of paraorthogonal
polynomials and those of orthogonal polynomials are very close. We
conjecture that these zeros are pairwise exponentially close with
the exception of $O((\ln N)^2)$ of them. We expect that the
distribution of the arguments of the zeros of orthogonal polynomials
on the unit circle is also Poisson.

3. We would also like to mention the related work of Bourget,
Howland, and Joye \cite{BHJ} and of Joye \cite{Joye2}, \cite{Joye3}.
In these papers, the authors analyze the spectral properties of a
class of five-diagonal random unitary matrices similar to the CMV
matrices (with the difference that it contains an extra random
parameter). In \cite{Joye3} (a preprint which appeared as this work
was being completed), the author considers a subclass of the class
of \cite{BHJ} that does not overlap with the orthogonal polynomials
on the unit circle and proves Aizenman-Molchanov bounds similar to
the ones we have in Section \ref{s3}.

4. The results presented in our paper were announced by Simon in
\cite{SPrep}, where he describes the distribution of the zeros of
orthogonal polynomials on the unit circle in two distinct (and, in a
certain way, opposite) situations. In the first case, of random
Verblunsky coefficients, our paper shows that there is no local
correlation between the zeros (Poisson behavior). The second case
consists of Verblusky coefficients given by the formula $\alpha_n =
C b^n + O( (b \Delta)^n)$ (i.e., $\alpha_n / b^n$ converges to a
constant $C$ sufficiently fast). In this case it is shown in
\cite{SPrep} that the zeros of the orthogonal polynomials are
equally spaced on the circle of radius $b$, which is equivalent to
saying that the angular distance between nearby zeros is $2 \pi / n$
(``clock" behavior).

\section*{Acknowledgements}
I am very grateful to B. Simon for his help and encouragement in the
investigation of this problem. I also want to thank M. Aizenman for
the personal communication \cite{AizPC}.


\end{document}